\documentclass[dvips]{article}
\usepackage{supertabular,lscape,epsfig}
\usepackage{amssymb}
\usepackage{amsmath}

\DeclareSymbolFont{ppa}{OT1}{ppl}{m}{it}
\DeclareMathSymbol{\vv}{\mathalpha}{ppa}{'166}

\begin{document}

\newcommand{\dd}{\,{\rm d}}
\newcommand{\ie}{{\it i.e.},\,}
\newcommand{\etal}{{\it et al.\ }}
\newcommand{\eg}{{\it e.g.},\,}
\newcommand{\cf}{{\it cf.\ }}
\newcommand{\vs}{{\it vs.\ }}
\newcommand{\zdot}{\makebox[0pt][l]{.}}
\newcommand{\up}[1]{\ifmmode^{\rm #1}\else$^{\rm #1}$\fi}
\newcommand{\dn}[1]{\ifmmode_{\rm #1}\else$_{\rm #1}$\fi}
\newcommand{\upd}{\up{d}}
\newcommand{\uph}{\up{h}}
\newcommand{\upm}{\up{m}}
\newcommand{\ups}{\up{s}}
\newcommand{\arcd}{\ifmmode^{\circ}\else$^{\circ}$\fi}
\newcommand{\arcm}{\ifmmode{'}\else$'$\fi}
\newcommand{\arcs}{\ifmmode{''}\else$''$\fi}
\newcommand{\MS}{{\rm M}\ifmmode_{\odot}\else$_{\odot}$\fi}
\newcommand{\RS}{{\rm R}\ifmmode_{\odot}\else$_{\odot}$\fi}
\newcommand{\LS}{{\rm L}\ifmmode_{\odot}\else$_{\odot}$\fi}

\newcommand{\Abstract}[2]{{\footnotesize\begin{center}ABSTRACT\end{center}
\vspace{1mm}\par#1\par
\noindent
{~}{\it #2}}}

\newcommand{\TabCap}[2]{\begin{center}\parbox[t]{#1}{\begin{center}
  \small {\spaceskip 2pt plus 1pt minus 1pt T a b l e}
  \refstepcounter{table}\thetable \\[2mm]
  \footnotesize #2 \end{center}}\end{center}}

\newcommand{\TableSep}[2]{\begin{table}[p]\vspace{#1}
\TabCap{#2}\end{table}}

\newcommand{\FigCap}[1]{\footnotesize\par\noindent Fig.\  %
  \refstepcounter{figure}\thefigure. #1\par}

\newcommand{\TableFont}{\footnotesize}
\newcommand{\TableFontIt}{\ttit}
\newcommand{\SetTableFont}[1]{\renewcommand{\TableFont}{#1}}

\newcommand{\MakeTable}[4]{\begin{table}[htb]\TabCap{#2}{#3}
  \begin{center} \TableFont \begin{tabular}{#1} #4 
  \end{tabular}\end{center}\end{table}}

\newcommand{\MakeTableSep}[4]{\begin{table}[p]\TabCap{#2}{#3}
  \begin{center} \TableFont \begin{tabular}{#1} #4 
  \end{tabular}\end{center}\end{table}}

\newcommand{\TabCapp}[2]{\begin{center}\parbox[t]{#1}{\centerline{
  \small {\spaceskip 2pt plus 1pt minus 1pt T a b l e}
  \refstepcounter{table}\thetable}
  \vskip0mm
  \centerline{\footnotesize #2}}
  \vskip0mm
\end{center}}

\newcommand{\MakeTableSepp}[4]{\begin{table}[p]\TabCapp{#2}{#3}
  \begin{center} \TableFont \begin{tabular}{#1} #4
  \end{tabular}\end{center}\end{table}}

\newenvironment{references}%
{
\footnotesize \frenchspacing
\renewcommand{\thesection}{}
\renewcommand{\in}{{\rm in }}
\renewcommand{\AA}{Astron.\ Astrophys.}
\newcommand{\AAS}{Astron.~Astrophys.~Suppl.~Ser.}
\newcommand{\ApJ}{Astrophys.\ J.}
\newcommand{\ApJS}{Astrophys.\ J.~Suppl.~Ser.}
\newcommand{\ApJL}{Astrophys.\ J.~Letters}
\newcommand{\AJ}{Astron.\ J.}
\newcommand{\IBVS}{IBVS}
\newcommand{\PASP}{P.A.S.P.}
\newcommand{\Acta}{Acta Astron.}
\newcommand{\MNRAS}{MNRAS}
\renewcommand{\and}{{\rm and }}
\section{{\rm REFERENCES}}
\sloppy \hyphenpenalty10000
\begin{list}{}{\leftmargin1cm\listparindent-1cm
\itemindent\listparindent\parsep0pt\itemsep0pt}}%
{\end{list}\vspace{2mm}}

\def\TYLDA{~}
\newlength{\DW}
\settowidth{\DW}{0}
\newcommand{\dw}{\hspace{\DW}}

\newcommand{\refitem}[5]{\item[]{#1} #2%
\def\REFARG{#3}\ifx\REFARG\TYLDA\else, {\it#3}\fi
\def\REFARG{#4}\ifx\REFARG\TYLDA\else, {\bf#4}\fi
\def\REFARG{#5}\ifx\REFARG\TYLDA\else, {#5}\fi.}

\newcommand{\Section}[1]{\section{\hskip-6mm.\hskip3mm#1}}
\newcommand{\Subsection}[1]{\subsection{#1}}
\newcommand{\Acknow}[1]{\par\vspace{5mm}{\bf Acknowledgements.} #1}
\pagestyle{myheadings}

\newfont{\bb}{ptmbi8t at 12pt}
\newcommand{\xrule}{\rule{0pt}{2.5ex}}
\newcommand{\xxrule}{\rule[-1.8ex]{0pt}{4.5ex}}
\def\thefootnote{\fnsymbol{footnote}}
\begin{center}
{\Large\bf The Optical Gravitational Lensing Experiment.
\vskip3pt
Eclipsing Binary Stars in the 
\vskip5pt
Large Magellanic Cloud\footnote{Based on  
observations obtained with the 1.3~m Warsaw telescope at the Las Campanas  
Observatory of the Carnegie Institution of Washington.}} 
\vskip 3mm
{\bf {\L}.~~W~y~r~z~y~k~o~w~s~k~i$^1$, ~~A.~~U~d~a~l~s~k~i$^1$, ~~M.
~~K~u~b~i~a~k$^1$, ~~M.~~S~z~y~m~a~{\'n}~s~k~i$^1$,~~
K.~~{\.Z}~e~b~r~u~{\'n}$^1$, ~~I.~~S~o~s~z~y~{\'n}~s~k~i$^1$,\\
P.R.~~W~o~{\'z}~n~i~a~k$^2$, ~~G.~~P~i~e~t~r~z~y~{\'n}~s~k~i$^{1,3}$~~and 
~~O.~~S~z~e~w~c~z~y~k$^1$}
\vskip 2mm
{$^1$ Warsaw University Observatory, Al.~Ujazdowskie~4, 00-478~Warsaw, Poland\\
e-mail: (wyrzykow,udalski,mk,msz,zebrun,soszynsk,pietrzyn,szewczyk)@astrouw.edu.pl\\
$^2$ Princeton University Observatory, Princeton, NJ 08544-1001, USA\\
Los Alamos National Observatory, MS-D436, Los Alamos NM 85745, USA\\
email: wozniak@lanl.gov\\
$^3$ Universidad de Concepci{\'o}n, Departamento de Fisica,
Casilla 160-C, Concepci{\'o}n, Chile\\
email: pietrzyn@hubble.cfm.udec.cl}
\end{center}

\vspace*{9pt}
\Abstract{We present the catalog of 2580 eclipsing binary stars detected in 
4.6 square degree area of the central parts of the Large Magellanic Cloud. The 
photometric data were collected during the second phase of the OGLE 
microlensing search from 1997 to 2000. The eclipsing objects were selected 
with the automatic search algorithm based on an artificial neural network. 
Basic statistics of eclipsing stars are presented. Also, the list of 36 
candidates of detached eclipsing binaries for spectroscopic study and for 
precise LMC distance determination is provided. The full catalog is accessible 
from the OGLE {\sc Internet} archive.}{binaries: eclipsing -- Magellanic 
Clouds -- Catalogs} 

\Section{Introduction}
Eclipsing binary stars are among the most important sources of information on 
stellar parameters like radii, masses, luminosities, etc. They also seem to be 
very promising candidates for standard candles. They should allow to determine 
distances within the Local Group with accuracy of a few percent (Paczy{\'n}ski 
1997). The method of distance determination to eclipsing systems is almost 
hundred years old and its main advantage is that it is largely geometric, \ie 
free from possible population effects which affect other standard candles. 
With accurate photometry and spectroscopy, absolute dimensions of both 
components can be precisely derived and, with accurately determined 
temperatures, emitted fluxes can be calculated. When compared with the flux 
observed from the Earth it should give precise distance to the binary. 

Eclipsing binary stars are very common in the Universe, but their detection 
may be quite difficult, because very good sampling of the light curve is 
necessary to detect eclipses. Large photometric databases collected during 
microlensing surveys provide 
ideal observational material for search for eclipsing objects. Hundreds of 
observations of millions of stars make detection of eclipsing objects 
relatively easy and efficient. For example, large samples of eclipsing binary 
stars were already found in  the Large Magellanic Cloud (Grison \etal 
1995, Alcock \etal 1997), Small Magellanic Cloud (Udalski \etal 1998a) or 
Galactic bulge (Udalski \etal 1997a). 

Determination of accurate distances is one of the most important goals of 
the modern astrophysics, in particular, the distance to the LMC as the 
extragalactic distance scale is based on the LMC distance. In the long 
lasting dispute on the LMC distance, large sample of eclipsing binary stars 
can potentially solve this problem. To date several applications of eclipsing 
binaries for distance determination to the LMC were presented: HV982 
(Fitzpatrick \etal 2002), EROS1044 (Ribas \etal 2002). However single stars 
determination can be affected by all kind of systematic errors (\eg HV2274: 
Guinan \etal 1998, Udalski \etal 1998b, Nelson \etal 2000, Groenewegen and 
Salaris 2001). Large and consistent sample of eclipsing stars is necessary for obtaining 
reliable distances to objects, as it was shown in Harries \etal (2003), based 
on OGLE-II catalog of about 1400 eclipsing binaries in the SMC (Udalski \etal 
1998a). 

The main aim of this paper is to provide a list and photometry of eclipsing 
binary stars detected in the Large Magellanic Cloud during the OGLE-II survey (Udalski, Kubiak and 
Szyma{\'n}ski 1997b). The catalog contains 2580 
objects found in the Difference Image Analysis Catalog of variable stars in 
the LMC ({\.Z}ebru{\'n} \etal 2001b). 

The detected eclipsing binaries were divided into three classical types of eclipsing variables: EA 
(Algol type), EB ($\beta$~Lyr type) and EW (W~~UMa type). The sample is 
reasonably complete, allowing statistical analysis and should provide a good 
material for testing theory of evolution of binary system as well as studying 
the LMC evolution, star formation etc. From detached systems, \ie EA class of 
eclipsing binaries, we additionally selected 36 objects -- the best candidates 
for distance determination to the LMC. We also describe automated algorithm 
based on artificial neural network developed for eclipsing star search which 
can be applied in the future for other large scale variable stars' 
classifications. 
 
\Section{Observational Data}
All photometric data presented in the catalog of eclipsing stars were 
collected with the 1.3-m Warsaw telescope at the Las Campanas Observatory, 
Chile, which is operated by the Carnegie Institution of Washington, during the 
second phase of the OGLE experiment. The telescope was equipped with the 
``first generation'' camera with the SITe 2048x2048 CCD detector working in 
driftscan mode. The pixel size was 24${\mu}$m giving the scale of 0.417 
arcsec/pixel. 

Observations of the LMC were performed in the ``slow'' reading mode of the CCD 
detector with the gain 3.8e\( ^{-} \)/ADU and readout noise of about 
5.4~e$^-$. Details of the instrumentation setup can be found in Udalski, 
Kubiak and Szyma{\'n}ski (1997b). 

Regular observations of the LMC fields started on January 6, 1997 and covered 
about 4.5 square degrees of the central parts of the LMC. Reductions of the 
photometric data collected up to the end of May 2000 were performed with the 
Difference Image Analysis (DIA) package (Wo{\'z}niak 2000, {\.Z}ebru{\'n}, 
Soszy{\'n}ski and Wo{\'z}niak 2001a) and variable stars candidates were 
published in the Catalog of variable stars in the Magellanic Clouds 
({\.Z}ebru{\'n} \etal 2001b). 

\MakeTable{rrr}{11cm}{Equatorial coordinates of the LMC fields}{
\hline
\multicolumn{1}{c}{Field}&\multicolumn{1}{c}{RA (J2000)}&
\multicolumn{1}{c}{DEC (J2000)}\\
\hline
LMC\_SC1  & 5\uph33\upm49\ups & $-70\arcd06\arcm10\arcs$\\
LMC\_SC2  & 5\uph31\upm17\ups & $-69\arcd51\arcm55\arcs$\\
LMC\_SC3  & 5\uph28\upm48\ups & $-69\arcd51\arcm55\arcs$\\
LMC\_SC4  & 5\uph26\upm18\ups & $-69\arcd48\arcm05\arcs$\\
LMC\_SC5  & 5\uph24\upm48\ups & $-69\arcd41\arcm05\arcs$\\
LMC\_SC6  & 5\uph21\upm18\ups & $-69\arcd37\arcm10\arcs$\\
LMC\_SC7  & 5\uph18\upm48\ups & $-69\arcd24\arcm10\arcs$\\
LMC\_SC8  & 5\uph16\upm18\ups & $-69\arcd19\arcm15\arcs$\\
LMC\_SC9  & 5\uph13\upm48\ups & $-69\arcd14\arcm05\arcs$\\
LMC\_SC10 & 5\uph11\upm16\ups & $-69\arcd09\arcm15\arcs$\\
LMC\_SC11 & 5\uph08\upm41\ups & $-69\arcd10\arcm05\arcs$\\
LMC\_SC12 & 5\uph06\upm16\ups & $-69\arcd38\arcm20\arcs$\\
LMC\_SC13 & 5\uph06\upm14\ups & $-68\arcd43\arcm30\arcs$\\
LMC\_SC14 & 5\uph03\upm49\ups & $-69\arcd04\arcm45\arcs$\\
LMC\_SC15 & 5\uph01\upm17\ups & $-69\arcd04\arcm45\arcs$\\
LMC\_SC16 & 5\uph36\upm18\ups & $-70\arcd09\arcm40\arcs$\\
LMC\_SC17 & 5\uph38\upm48\ups & $-70\arcd16\arcm45\arcs$\\
LMC\_SC18 & 5\uph41\upm18\ups & $-70\arcd24\arcm50\arcs$\\
LMC\_SC19 & 5\uph43\upm48\ups & $-70\arcd34\arcm45\arcs$\\
LMC\_SC20 & 5\uph46\upm18\ups & $-70\arcd44\arcm50\arcs$\\
LMC\_SC21 & 5\uph21\upm14\ups & $-70\arcd33\arcm20\arcs$\\
\hline}
The DIA photometry is based on the {\it I}-band observations. The catalog of 
variable stars contains about 53\,000 stars in 21 fields of the LMC (Table~1). 
Each star has at least 300 photometric measurements. The magnitudes of stars 
were transformed to the standard system (Udalski \etal 2000). The errors of 
the measurements are about 0.005 mag for the brightest stars (${I<16}$~mag) 
and grow to 0.08~mag at 19~mag and to 0.3~mag at 20.5~mag. 
\vskip9mm
\Section{Search for Eclipsing Binary Stars}
The most common method of classification of variable stars is based on visual 
inspection of light curves, but when the number of stars to be examined is 
growing to thousands, it becomes very inefficient. Until now several attempts 
have been made to create an automated periodic variables classification. First 
method is examination of two dimensional projections of a multidimensional 
parameter space. Such an approach was applied by \eg Ruci{\'n}ski (1993, 
1997), Szyma{\'n}ski, Kubiak, Udalski (2001), for contact binaries, Mizerski 
and Bejger (2001) for RR~Lyr and W~UMa stars, Udalski \etal (1999) for 
Cepheids. Expanded and improved versions of this method were also applied by 
all sky surveys, as ROTSE (Akerlof 2000) and ASAS (Pojma{\'n}ski 2002) to 
classify variables into several common variability types. 

Another method applied in automated classifications includes neural networks 
and machine learning algorithms (\eg Wo{\'z}niak \etal 2001). We attempted to 
use an artificial neural network as a main classification tool to find 
eclipsing binaries among OGLE-II variable stars in the LMC. 
\vskip4mm
\Subsection{Preparation of Photometric Data}
We performed the search on about 53000 stars from 21 fields of the LMC, 
published in the OGLE-II catalog of variable stars ({\.Z}ebru{\'n} \etal 
2001b). 

First, all stars from the DIA catalog were checked for all kind irregularities in 
the light curve. This step allowed us to reject most non-periodic variables from the 
database. Next, all stars left (about 36\,000) were searched for periodicity 
using {\sc AoV} algorithm (Schwarzenberg-Czerny 1989). Because photometric 
data span about 1500 days we searched for periodicities in the wide range of  
0.1--500~days. 

Before the main process of recognition of variability types was started, some 
additional preparation steps were performed. To use the neural network, we had 
to convert the light curves of all periodic stars in such a way that the differences between 
them were only caused by differences of variability type. The main problem was 
caused by the {\sc AoV} algorithm which sometimes detects not the correct period, $P$, 
but ${2\times P}$ or $P/2$. Therefore we performed a Fourier decomposition of 
phased light curves and examined the ratio of the first two coefficients. We did 
not change detected period if the ratio indicated that the first harmonic 
dominates, and divided it by two in other cases. In this way we further 
processed all eclipsing variables phased with periods $P/2$, and all pulsating 
(sinusoidal and wide class of ``saw shape'' type) with their correct periods. 

At the last stage before running the network recognition we found zero phase 
for each light curve and we mapped the light curve as a ${70\times15}$ pixel 
image. Examples of the projection of light curves are shown in Fig.~1.
\begin{figure}[htb]
\hglue-8mm{\includegraphics[width=7cm]{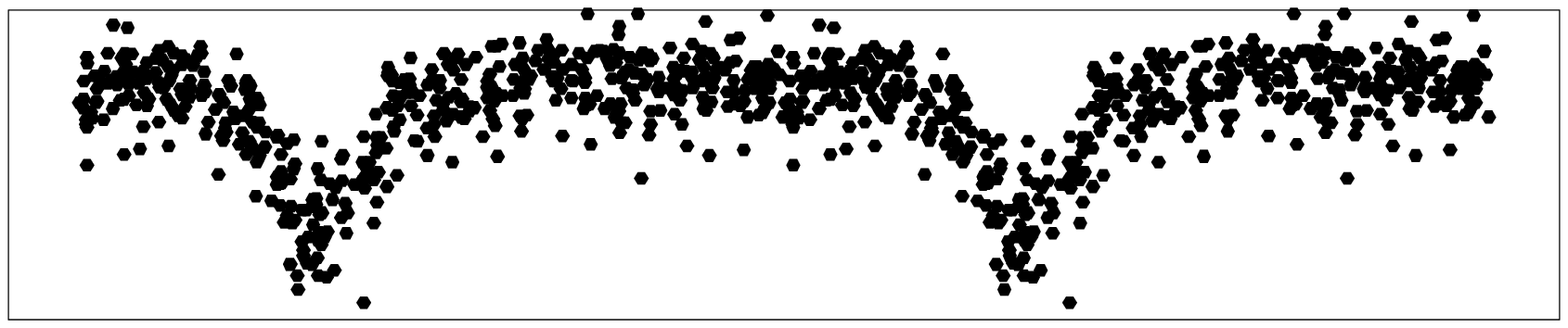}\hskip-7mm\includegraphics[width=7cm]{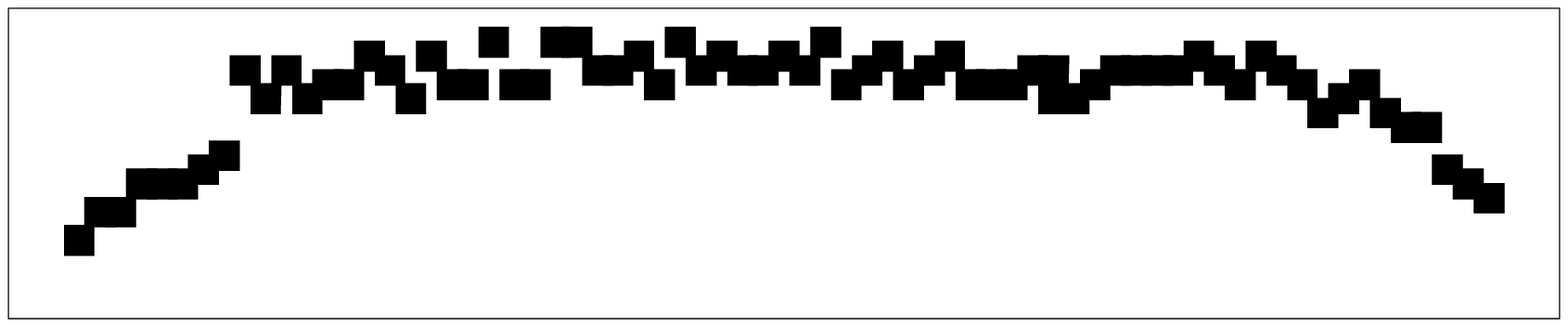}} 
\hglue-8mm{\includegraphics[width=7cm]{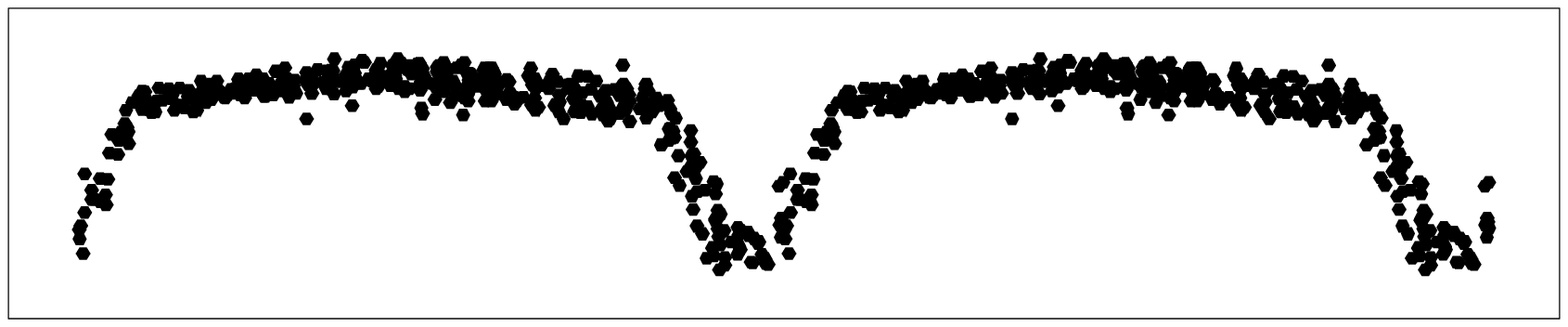}\hskip-7mm\includegraphics[width=7cm]{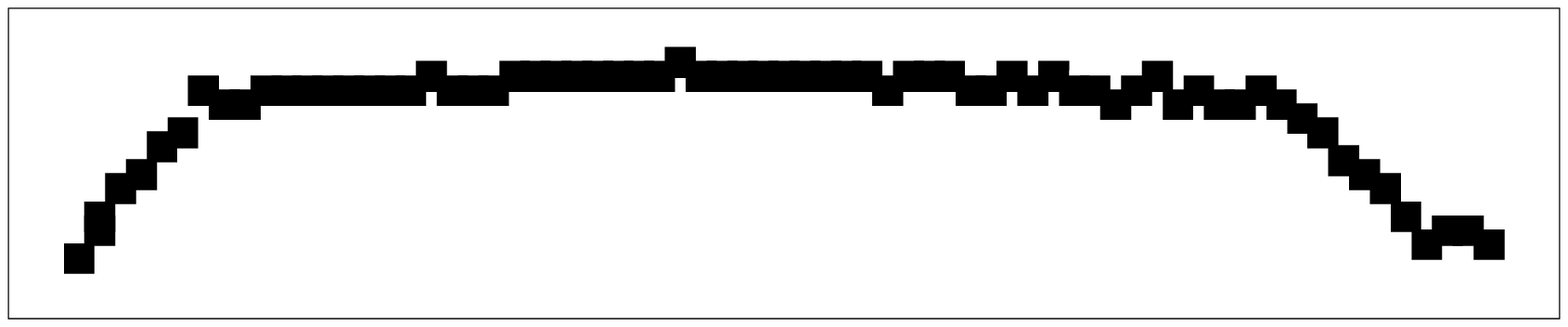}} 
\hglue-8mm{\includegraphics[width=7cm]{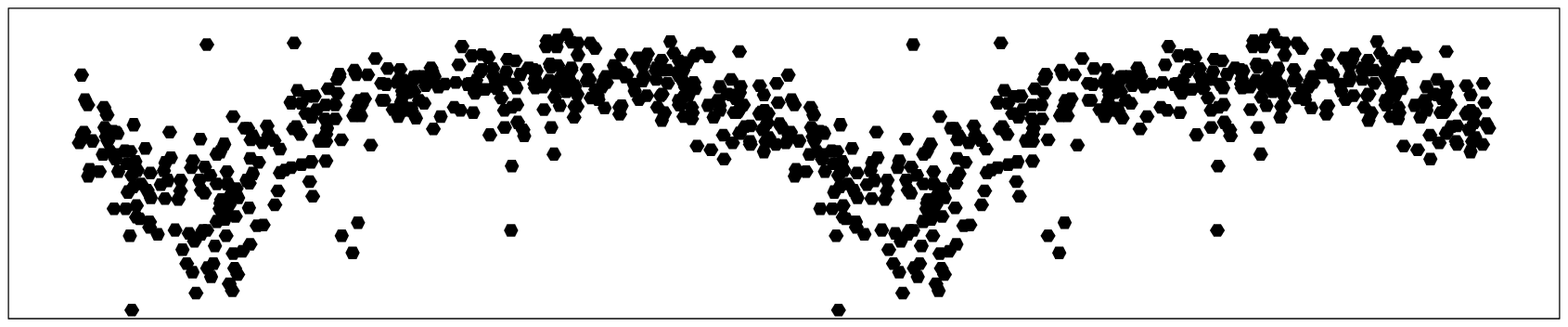}\hskip-7mm\includegraphics[width=7cm]{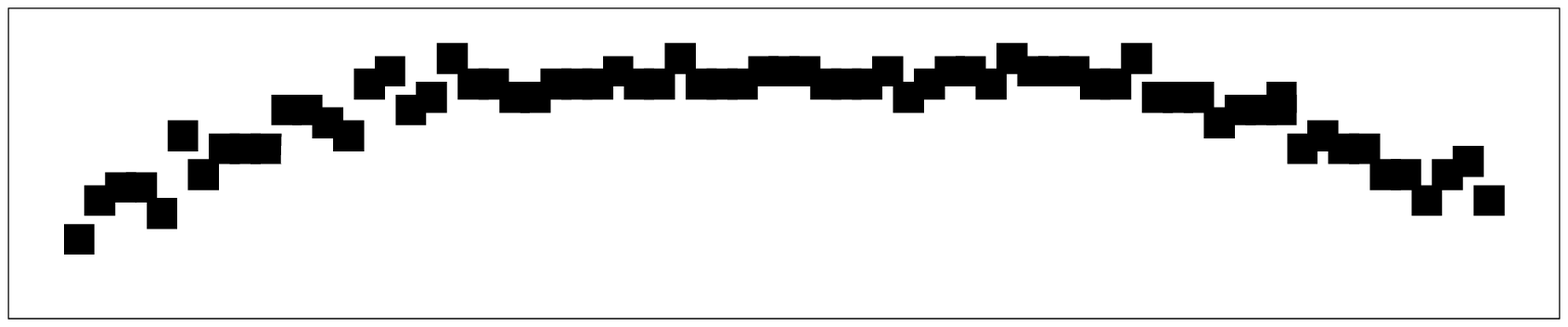}} 
\hglue-8mm{\includegraphics[width=7cm]{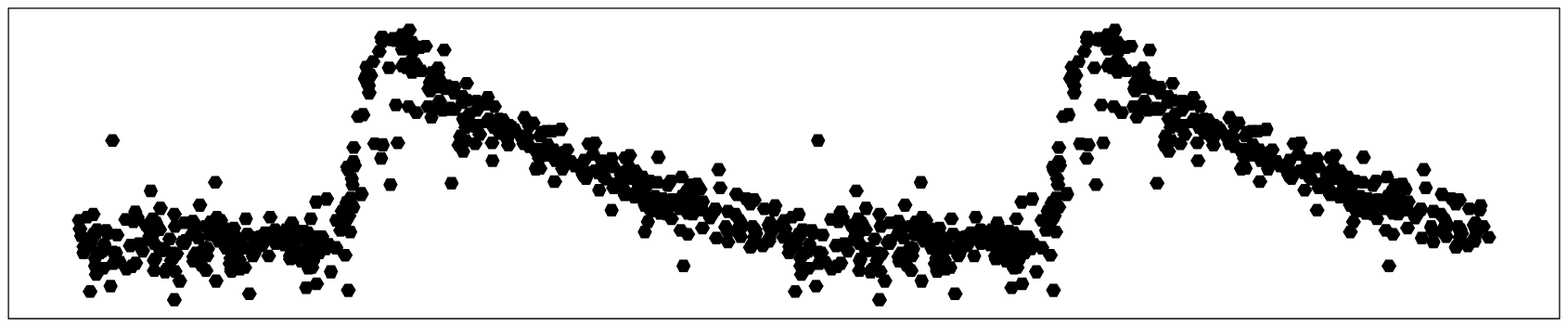}\hskip-7mm\includegraphics[width=7cm]{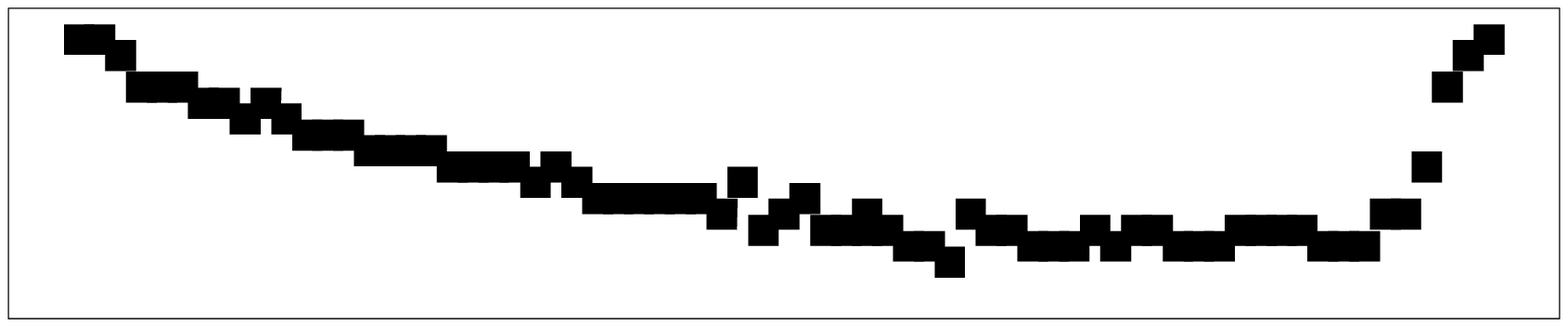}} 
\FigCap{Examples of conversion of phased light curves {\it (left)} to 
${70\times15}$ pixel images {\it (right)}, which then entered the neural 
network. Notice that the light curves are shown twice (phases 0--2) for 
clarity, but the maps are shown only once, as they entered the network.} 
\end{figure}

\Subsection{Neural Network}
Now, the variability type recognition problem becomes the image recognition 
problem, which we solved using artificial neural network. Fig.~2 shows 
schematically the structure of the network. We built three layers, 
one-direction, non-linear neural network with logistic activation function. 
Input of the network was a ${70\times15}$ pixel image of the light curve, 
which was transformed to one-dimensional line-by-line ``image''. Therefore we 
had 1050 input nodes in the Input Layer. Layers I, II and III (Output Layer) 
consisted of 150, 50 and 5 neurons, respectively. Number of neurons in each 
layer was chosen accordingly to the image recognition requirements. 
\begin{figure}[htb]
\vspace{3cm}
\centerline{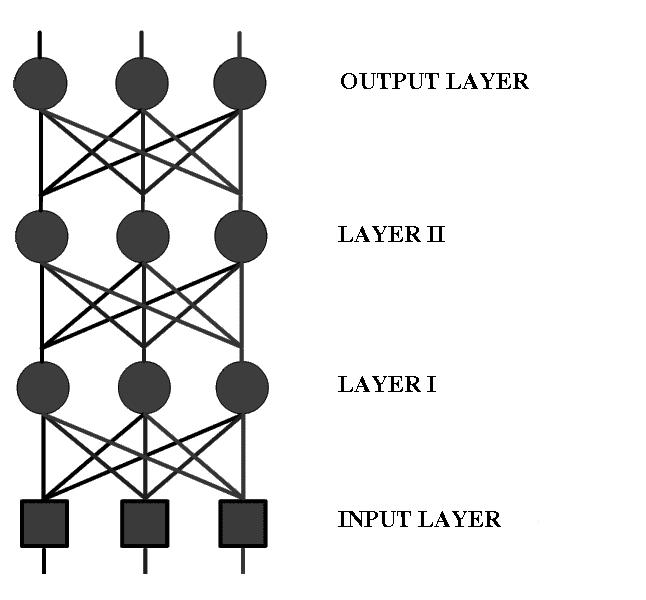}
\vspace{3.5cm}
\FigCap{Schematic structure of artificial neural network used for automated 
search of eclipsing binary stars. Input layer has 1050 nodes, Layer~I, II and 
Output Layer have 150, 50 and 5 neurons, respectively.}
\end{figure} 

We used error back-propagation algorithm for network learning. Neural network 
was constructed to recognize three main types of variable stars, according to 
their light curve shape: eclipsing, sinusoidal and saw-shape type. From the 
first LMC field (LMC\_SC1) we selected manually 10 examples of each type 
objects and presented them randomly to the network a few thousand times, until 
the mean network error was smaller than $10^{-8}$. 

After learning, light curves of all previously found periodical stars from all 
LMC fields were subject of the network analysis. The procedure allowed us to 
divide stars into main predefined types and exclude some artifacts and non-variable 
objects from the catalog. After this stage we could exclude the sinusoidal and 
saw-shaped variables and we were left with only about 3000 candidates for 
eclipsing variables. 

\Subsection{Detailed Classification of Eclipsing Variable Stars}
In order to divide the obtained database of the eclipsing stars into 
subclasses we inspected visually light curves of all candidates. First, 
periods of eclipsing stars used in the network 
analysis were multiplied by two to obtain real periods. Then, the 
periods were tuned up to smooth the eclipse shape which is very sensitive to 
period inaccuracies. Similarly to the 
Fourth Edition of ``General Catalog of Variable Stars'' (GCVS: Kholopov \etal 
1999) we divided detected eclipsing variables into three classical types, 
based on the shape of the light curve: EA (Algol type), EB ($\beta$~Lyr type) 
and EW (W~UMa type). For several stars dual classification (\eg EB/EW) was 
chosen, because of difficulties with distinguishing between two classes. The 
most difficult classification problem was the separation between EB and EW 
classes. When the orbital period was about one day or less we usually assigned 
EW type. In the case of several stars their variability, classification or 
period are uncertain. Such objects are marked with additional remark as 
``uncertain''. 

By the  visual inspection we excluded very uncertain objects and almost 
200 light curves of probably ellipsoidal variables -- stars with very low 
amplitude (about 0.1~mag) and in very wide range of periods. They were 
originally classified as eclipsing stars because the shape of their light 
curves revealed somewhat different depths of minima. The remaining part of 
ellipsoidal variables with clearly sinusoidal light curves were included in 
the ``sinusoidal shape'' class during neural network classification. However, 
in the case of some stars we were still unable to clearly distinguish between 
eclipsing and ellipsoidal variables. Therefore, they were also marked  
as ELL. 

In the case of some objects additional variability of one or both components 
was superimposed on the clear eclipsing variability. These light variations 
could be caused by \eg spots on binary system stars, high proper motion of the 
system (what is seen in the DIA method as a long term falling or rising 
tendency in the light curve, see Soszy{\'n}ski \etal 2002) or probably by 
pulsations of one of the binary components. The latter type systems containing 
Cepheids were already found in the LMC by OGLE (Udalski \etal 1999) and 
examined by MACHO Collaboration (Alcock \etal 2002). All variables with 
additional, confirmed or only suspected, light curve changes  are marked with 
``Puls'' or ``Puls?'' remark, respectively. 

Additionally, we found 291 eclipsing variables with clear eccentricity effects 
visible in their light curves. They were marked with ``ecc''. In 20 cases we 
could not smooth both eclipses using the same period what suggests large 
appsidal motion. We marked these objects as ``eccAP'' and selected the period 
corresponding to the primary minimum. 
\vspace*{-9pt}
\Section{Catalog of Eclipsing Binary Stars}
\vspace*{-3pt}
In total 2580 eclipsing binary stars were found in the OGLE-II DIA catalog of 
variable stars in the LMC fields. List of the first 50 stars is presented in 
Table~2. It contains the ordinal number of the eclipsing variable star, field, 
name of the star, orbital period, heliocentric Julian Date of the primary 
minimum ($T_0$ -- 2\,450\,000, corrected for position of the star in the 
driftscan image, as described by {\.Z}ebru{\'n} \etal 2001b), {\it V}-band 
magnitude, ${B-V}$ and ${V-I}$ colors at maximum brightness from the standard 
OGLE-II data pipeline photometry, amplitude in the {\it I}-band from DIA 
photometry (depth of primary minimum) and eclipsing type. Color value of 
${-99.99}$ stands for no observations in the {\it B} or {\it V} bands.
\renewcommand{\TableFont}{\scriptsize}
\renewcommand{\arraystretch}{1.05}
\MakeTableSep{r
@{\hspace{5pt}}r
@{\hspace{5pt}}r
@{\hspace{2pt}}r
@{\hspace{3pt}}r
@{\hspace{4pt}}r
@{\hspace{1pt}}r
@{\hspace{3pt}}r
@{\hspace{3pt}}r
@{\hspace{3pt}}r}{11cm}{Eclipsing binaries in the LMC}{
\hline
\multicolumn{1}{c}{No.}&
\multicolumn{1}{c}{Field}&
\multicolumn{1}{c}{Star}&
\multicolumn{1}{c}{~Period}&
\multicolumn{1}{c}{$T_0$}&
\multicolumn{4}{c}{$V$~~~ $B{-}V$~~ $V{-}I$~$\Delta I_{\rm PRI}$}&
\multicolumn{1}{c}{~~Type}\\
&&&
\multicolumn{1}{c}{~[days]}&
\multicolumn{1}{c}{--2450000}&
&
&
\multicolumn{2}{r}{(DIA)}
&\\
\noalign{\vskip3pt}
\hline
1 &LMC\_SC1&OGLE053227.36-701148.4&  1.843800&455.45870&19.16&$-99.99$&$ 0.49$&0.61&EB/EW\\
2 &LMC\_SC1&OGLE053230.79-700838.3&  3.696430&455.45331&18.68&$  0.34$&$ 0.49$&0.98&EA\\
3 &LMC\_SC1&OGLE053232.28-700056.5&  1.225200&454.01971&17.69&$ -0.05$&$ 0.11$&0.25&EB\\
4 &LMC\_SC1&OGLE053234.48-700218.2&  1.577020&455.67840&18.33&$  0.06$&$ 0.17$&0.44&EA\\
5 &LMC\_SC1&OGLE053235.27-702606.7&  1.663420&456.09860&17.44&$ -0.04$&$ 0.05$&0.19&EB\\
6 &LMC\_SC1&OGLE053236.25-700644.5&  2.659240&457.37322&17.38&$ -0.16$&$ 0.11$&0.22&EA\\
7 &LMC\_SC1&OGLE053236.80-695405.7&  3.617860&452.71699&17.28&$ -0.07$&$-0.01$&0.26&EA\\
8 &LMC\_SC1&OGLE053236.97-700317.1&  1.849470&454.22510&19.01&$ -0.01$&$ 0.22$&1.11&EB\\
9 &LMC\_SC1&OGLE053238.66-701354.2&  5.481640&455.44523&17.47&$ -0.08$&$-0.05$&0.27&EA\\
10&LMC\_SC1&OGLE053240.31-701127.8&  1.905630&456.59061&18.46&$ -0.04$&$ 0.25$&1.45&EA\\
11&LMC\_SC1&OGLE053241.43-694149.3&136.423800&491.06174&18.55&$  1.04$&$ 1.19$&0.15&EA\\
12&LMC\_SC1&OGLE053241.93-695109.2&  4.954340&453.59493&16.87&$ -0.12$&$-0.04$&0.54&EA\\
13&LMC\_SC1&OGLE053242.83-700513.5&  0.958740&455.71161&18.19&$ -0.31$&$ 0.40$&0.31&EA\\
14&LMC\_SC1&OGLE053244.07-695406.8&  2.314920&455.76774&19.02&$  0.06$&$ 0.39$&0.47&EA\\
15&LMC\_SC1&OGLE053244.79-701621.5&  3.047280&452.04025&19.92&$-99.99$&$ 0.78$&1.31&EA\\
16&LMC\_SC1&OGLE053247.54-694403.1&  0.331035&454.77797&17.34&$  0.63$&$ 0.89$&0.42&EW\\
17&LMC\_SC1&OGLE053248.27-703013.0&  1.115930&455.09168&19.01&$  0.03$&$ 0.17$&0.73&EA\\
18&LMC\_SC1&OGLE053249.77-695445.5&  5.749730&454.60472&16.94&$  0.00$&$ 0.08$&0.69&EA\\
19&LMC\_SC1&OGLE053251.37-701446.9&  6.162960&450.94134&19.04&$  0.12$&$ 0.49$&0.79&EA\\
20&LMC\_SC1&OGLE053251.73-700256.2&  0.297175&454.67085&17.71&$  0.86$&$ 0.98$&0.77&EW\\
21&LMC\_SC1&OGLE053252.71-694837.6&  6.964140&454.61819&19.41&$  1.05$&$ 1.01$&0.23&EA\\
22&LMC\_SC1&OGLE053255.08-695837.9&  0.901480&455.28013&18.86&$  0.19$&$ 0.22$&0.65&EA\\
23&LMC\_SC1&OGLE053257.90-700658.5&  1.289960&456.12106&18.84&$  0.14$&$ 0.25$&0.64&EA\\
24&LMC\_SC1&OGLE053259.41-701152.3&  5.738280&451.11831&19.59&$  0.24$&$ 0.61$&1.59&EA\\
25&LMC\_SC1&OGLE053259.97-695638.6&  3.368120&456.82108&17.41&$ -0.02$&$ 0.10$&1.59&EA\\
26&LMC\_SC1&OGLE053300.70-701934.1&  1.959150&455.52318&18.72&$  0.15$&$ 0.29$&0.95&EA\\
27&LMC\_SC1&OGLE053301.16-700805.4&  3.340010&451.84411&19.45&$  0.65$&$ 0.76$&0.59&EB\\
28&LMC\_SC1&OGLE053302.18-700237.1&  2.940790&457.21649&18.80&$  0.41$&$ 0.13$&0.82&EA\\
29&LMC\_SC1&OGLE053303.00-700516.8&  1.115260&455.67346&18.69&$ -0.08$&$ 0.02$&0.83&EA\\
30&LMC\_SC1&OGLE053304.23-701613.1& 69.773770&441.26483&17.79&$  0.79$&$ 0.94$&0.16&EA\\
31&LMC\_SC1&OGLE053306.15-695252.0&  5.487330&457.61490&19.85&$  0.22$&$ 0.61$&1.54&EA\\
32&LMC\_SC1&OGLE053308.15-700112.0&  0.877760&456.16928&18.58&$  0.07$&$ 0.11$&0.61&EB\\
33&LMC\_SC1&OGLE053308.85-695150.3&  1.313040&454.49957&19.46&$  0.13$&$ 0.33$&1.46&EA\\
34&LMC\_SC1&OGLE053310.22-694618.9& 26.451300&457.80026&18.88&$  0.89$&$ 1.09$&0.43&EB\\
35&LMC\_SC1&OGLE053311.25-702135.3&  5.225480&455.47968&19.79&$  0.39$&$ 0.42$&1.56&EA\\
36&LMC\_SC1&OGLE053311.63-694400.2&  2.144280&456.27341&18.65&$  0.01$&$ 0.04$&0.67&EA-ecc\\
37&LMC\_SC1&OGLE053311.89-694655.0&  6.194370&452.65093&19.52&$  0.10$&$ 0.48$&1.47&EA\\
38&LMC\_SC1&OGLE053312.11-701124.8&  7.937130&459.10728&18.56&$  0.13$&$ 0.18$&0.59&EA\\
39&LMC\_SC1&OGLE053312.82-700702.5&  5.394410&454.74615&17.18&$ -0.06$&$-0.02$&0.38&EA\\
40&LMC\_SC1&OGLE053314.45-701708.8&  2.580720&453.36055&18.46&$ -0.04$&$ 0.09$&0.80&EA\\
41&LMC\_SC1&OGLE053314.63-701005.3&  1.795680&456.15587&16.94&$  0.31$&$ 0.04$&0.18&EB/EW/ELL\\
42&LMC\_SC1&OGLE053317.37-695605.2& 23.913340&443.18485&19.47&$  0.89$&$ 1.41$&1.02&EA\\
43&LMC\_SC1&OGLE053318.69-695502.6&  4.637010&452.73410&16.17&$ -0.02$&$ 0.01$&0.90&EB\\
44&LMC\_SC1&OGLE053320.14-702025.0& 53.051330&525.69769&18.71&$  0.88$&$ 1.14$&3.55&EA\\
45&LMC\_SC1&OGLE053320.76-695427.7&  1.559150&456.41948&17.57&$ -0.12$&$ 0.02$&0.59&EA\\
46&LMC\_SC1&OGLE053321.09-701743.3&  6.108740&457.02623&19.52&$  0.64$&$ 1.17$&0.65&EA\\
47&LMC\_SC1&OGLE053323.38-700802.3&  7.187350&456.73297&16.94&$  0.00$&$ 0.16$&0.80&EA\\
48&LMC\_SC1&OGLE053326.25-703114.1&  2.945830&455.34469&17.53&$ -0.05$&$ 0.10$&0.43&EA\\
49&LMC\_SC1&OGLE053326.64-694404.2&  3.546780&454.60046&19.63&$  0.26$&$ 0.55$&1.18&EA\\
50&LMC\_SC1&OGLE053327.35-695205.5&  4.654350&451.62319&19.86&$  0.25$&$ 0.87$&1.37&EA\\
\hline}

One should remember that the conversion of the DIA flux differences to the 
magnitude scale is not always accurate. In particular, in the case of severely 
blended objects the depth of the maxima can be unreliable, as the constant 
flux cannot be accurately determined. Nevertheless, such blends contain a real 
eclipsing star. 

Among 2580 stars, 101 were identified twice in the overlapping regions between 
neighboring fields, so the total number of identified eclipsing binary stars 
is equal to 2681. List of all cross-identified objects is presented in Table~3. 
\MakeTableSep{l@{\hspace{3pt}}rl@{\hspace{3pt}}r}{11cm}{Cross-identification 
of eclipsing binary stars detected in overlapping regions}{
LMC\_SC1 $\leftrightarrow$ LMC\_SC16 & OGLE053458.41-701653.3&
LMC\_SC1 $\leftrightarrow$ LMC\_SC16 & OGLE053459.53-694406.2\\
LMC\_SC1 $\leftrightarrow$ LMC\_SC16 & OGLE053500.64-701843.0&
LMC\_SC1 $\leftrightarrow$ LMC\_SC16 & OGLE053502.18-694417.8\\
LMC\_SC1 $\leftrightarrow$ LMC\_SC16 & OGLE053504.85-695812.0&
LMC\_SC1 $\leftrightarrow$ LMC\_SC16 & OGLE053509.05-700659.9\\
LMC\_SC1 $\leftrightarrow$ LMC\_SC16 & OGLE053509.40-694631.1&
LMC\_SC1 $\leftrightarrow$ LMC\_SC16 & OGLE053512.36-702808.5\\
LMC\_SC1 $\leftrightarrow$ LMC\_SC2 & OGLE053227.36-701148.4&
LMC\_SC1 $\leftrightarrow$ LMC\_SC2 & OGLE053230.79-700838.3\\
LMC\_SC1 $\leftrightarrow$ LMC\_SC2 & OGLE053234.48-700218.2&
LMC\_SC1 $\leftrightarrow$ LMC\_SC2 & OGLE053236.80-695405.7\\
LMC\_SC1 $\leftrightarrow$ LMC\_SC2 & OGLE053236.97-700317.1&
LMC\_SC1 $\leftrightarrow$ LMC\_SC2 & OGLE053238.66-701354.2\\
LMC\_SC2 $\leftrightarrow$ LMC\_SC3 & OGLE052956.44-694220.9&
LMC\_SC2 $\leftrightarrow$ LMC\_SC3 & OGLE052957.18-701054.0\\
LMC\_SC2 $\leftrightarrow$ LMC\_SC3 & OGLE052957.34-695520.8&
LMC\_SC2 $\leftrightarrow$ LMC\_SC3 & OGLE052958.05-701224.3\\
LMC\_SC2 $\leftrightarrow$ LMC\_SC3 & OGLE052958.13-695936.2&
LMC\_SC2 $\leftrightarrow$ LMC\_SC3 & OGLE052958.54-695708.0\\
LMC\_SC2 $\leftrightarrow$ LMC\_SC3 & OGLE052959.80-694358.8&
LMC\_SC2 $\leftrightarrow$ LMC\_SC3 & OGLE052959.87-695950.1\\
LMC\_SC2 $\leftrightarrow$ LMC\_SC3 & OGLE052959.88-695918.6&
LMC\_SC2 $\leftrightarrow$ LMC\_SC3 & OGLE053000.05-695214.0\\
LMC\_SC2 $\leftrightarrow$ LMC\_SC3 & OGLE053001.64-693057.2&
LMC\_SC2 $\leftrightarrow$ LMC\_SC3 & OGLE053002.59-700945.8\\
LMC\_SC2 $\leftrightarrow$ LMC\_SC3 & OGLE053005.04-692750.0&
LMC\_SC2 $\leftrightarrow$ LMC\_SC3 & OGLE053005.21-694527.1\\
LMC\_SC2 $\leftrightarrow$ LMC\_SC3 & OGLE053006.98-695309.6&
LMC\_SC2 $\leftrightarrow$ LMC\_SC3 & OGLE053007.56-695321.3\\
LMC\_SC3 $\leftrightarrow$ LMC\_SC4 & OGLE052726.58-700720.2&
LMC\_SC3 $\leftrightarrow$ LMC\_SC4 & OGLE052728.35-694902.1\\
LMC\_SC3 $\leftrightarrow$ LMC\_SC4 & OGLE052728.57-694457.5&
LMC\_SC3 $\leftrightarrow$ LMC\_SC4 & OGLE052729.79-695936.0\\
LMC\_SC3 $\leftrightarrow$ LMC\_SC4 & OGLE052730.73-695244.0&
LMC\_SC3 $\leftrightarrow$ LMC\_SC4 & OGLE052736.26-693754.1\\
LMC\_SC3 $\leftrightarrow$ LMC\_SC4 & OGLE052737.48-692053.5&
LMC\_SC4 $\leftrightarrow$ LMC\_SC5 & OGLE052456.31-700250.4\\
LMC\_SC4 $\leftrightarrow$ LMC\_SC5 & OGLE052456.76-694324.9&
LMC\_SC4 $\leftrightarrow$ LMC\_SC5 & OGLE052456.78-694130.5\\
LMC\_SC4 $\leftrightarrow$ LMC\_SC5 & OGLE052458.44-693855.5&
LMC\_SC4 $\leftrightarrow$ LMC\_SC5 & OGLE052459.33-692607.2\\
LMC\_SC4 $\leftrightarrow$ LMC\_SC5 & OGLE052500.95-693616.2&
LMC\_SC4 $\leftrightarrow$ LMC\_SC5 & OGLE052501.42-695508.7\\
LMC\_SC4 $\leftrightarrow$ LMC\_SC5 & OGLE052501.51-700810.2&
LMC\_SC4 $\leftrightarrow$ LMC\_SC5 & OGLE052503.58-694030.2\\
LMC\_SC4 $\leftrightarrow$ LMC\_SC5 & OGLE052503.66-692849.5&
LMC\_SC4 $\leftrightarrow$ LMC\_SC5 & OGLE052503.68-692326.8\\
LMC\_SC4 $\leftrightarrow$ LMC\_SC5 & OGLE052504.39-695909.7&
LMC\_SC4 $\leftrightarrow$ LMC\_SC5 & OGLE052504.94-694849.3\\
LMC\_SC4 $\leftrightarrow$ LMC\_SC5 & OGLE052506.70-695640.8&
LMC\_SC4 $\leftrightarrow$ LMC\_SC5 & OGLE052509.46-700422.6\\
LMC\_SC5 $\leftrightarrow$ LMC\_SC6 & OGLE052229.91-691909.0&
LMC\_SC5 $\leftrightarrow$ LMC\_SC6 & OGLE052230.88-694409.1\\
LMC\_SC5 $\leftrightarrow$ LMC\_SC6 & OGLE052233.86-693256.4&
LMC\_SC5 $\leftrightarrow$ LMC\_SC6 & OGLE052235.40-693649.2\\
LMC\_SC5 $\leftrightarrow$ LMC\_SC6 & OGLE052235.87-695257.9&
LMC\_SC5 $\leftrightarrow$ LMC\_SC6 & OGLE052236.06-694026.6\\
LMC\_SC6 $\leftrightarrow$ LMC\_SC7 & OGLE051958.16-692823.9&
LMC\_SC6 $\leftrightarrow$ LMC\_SC7 & OGLE051959.79-692201.9\\
LMC\_SC6 $\leftrightarrow$ LMC\_SC7 & OGLE052002.15-692007.0&
LMC\_SC6 $\leftrightarrow$ LMC\_SC7 & OGLE052003.88-692616.1\\
LMC\_SC6 $\leftrightarrow$ LMC\_SC7 & OGLE052007.03-692925.6&
LMC\_SC6 $\leftrightarrow$ LMC\_SC7 & OGLE052008.00-693241.4\\
LMC\_SC7 $\leftrightarrow$ LMC\_SC8 & OGLE051729.24-692809.7&
LMC\_SC7 $\leftrightarrow$ LMC\_SC8 & OGLE051734.42-694422.2\\
LMC\_SC7 $\leftrightarrow$ LMC\_SC8 & OGLE051734.54-692736.5&
LMC\_SC7 $\leftrightarrow$ LMC\_SC8 & OGLE051736.06-693846.9\\
LMC\_SC7 $\leftrightarrow$ LMC\_SC8 & OGLE051737.60-693139.8&
LMC\_SC8 $\leftrightarrow$ LMC\_SC9 & OGLE051500.27-691126.5\\
LMC\_SC8 $\leftrightarrow$ LMC\_SC9 & OGLE051502.03-691918.0&
LMC\_SC8 $\leftrightarrow$ LMC\_SC9 & OGLE051502.88-690816.7\\
LMC\_SC8 $\leftrightarrow$ LMC\_SC9 & OGLE051506.39-693434.1&
LMC\_SC10 $\leftrightarrow$ LMC\_SC9 & OGLE051230.07-690624.2\\
LMC\_SC10 $\leftrightarrow$ LMC\_SC9 & OGLE051232.66-685637.9&
LMC\_SC11 $\leftrightarrow$ LMC\_SC12 & OGLE050726.38-693245.7\\
LMC\_SC11 $\leftrightarrow$ LMC\_SC12 & OGLE050728.63-692539.6&
LMC\_SC11 $\leftrightarrow$ LMC\_SC12 & OGLE050730.20-693638.8\\
LMC\_SC11 $\leftrightarrow$ LMC\_SC12 & OGLE050731.85-691010.9&
LMC\_SC11 $\leftrightarrow$ LMC\_SC13 & OGLE050731.61-690926.9\\
LMC\_SC12 $\leftrightarrow$ LMC\_SC13 & OGLE050511.63-691012.4&
LMC\_SC12 $\leftrightarrow$ LMC\_SC13 & OGLE050609.82-691109.9\\
LMC\_SC12 $\leftrightarrow$ LMC\_SC13 & OGLE050638.95-691026.0&
LMC\_SC12 $\leftrightarrow$ LMC\_SC13 & OGLE050659.47-691050.0\\
LMC\_SC12 $\leftrightarrow$ LMC\_SC13 & OGLE050712.75-691100.3&
LMC\_SC13 $\leftrightarrow$ LMC\_SC14 & OGLE050503.92-685221.0\\
LMC\_SC13 $\leftrightarrow$ LMC\_SC14 & OGLE050504.24-685758.9&
LMC\_SC13 $\leftrightarrow$ LMC\_SC14 & OGLE050504.90-690554.1\\
LMC\_SC14 $\leftrightarrow$ LMC\_SC15 & OGLE050234.59-690546.0&
LMC\_SC16 $\leftrightarrow$ LMC\_SC17 & OGLE053726.41-702422.8\\
LMC\_SC16 $\leftrightarrow$ LMC\_SC17 & OGLE053734.07-701253.7&
LMC\_SC16 $\leftrightarrow$ LMC\_SC17 & OGLE053734.55-694954.5\\
LMC\_SC16 $\leftrightarrow$ LMC\_SC17 & OGLE053738.73-700112.9&
LMC\_SC17 $\leftrightarrow$ LMC\_SC18 & OGLE054003.85-703837.3\\
LMC\_SC17 $\leftrightarrow$ LMC\_SC18 & OGLE054008.14-701202.8&
LMC\_SC17 $\leftrightarrow$ LMC\_SC18 & OGLE054008.22-700359.2\\
LMC\_SC17 $\leftrightarrow$ LMC\_SC18 & OGLE054009.08-703718.2&
LMC\_SC17 $\leftrightarrow$ LMC\_SC18 & OGLE054010.85-703013.4\\
LMC\_SC18 $\leftrightarrow$ LMC\_SC19 & OGLE054229.84-701154.9&
LMC\_SC19 $\leftrightarrow$ LMC\_SC20 & OGLE054509.85-703644.7\\
LMC\_SC21 $\leftrightarrow$ LMC\_SC5~~~& OGLE052230.33-700651.6\\
}

Stars' names follow the convention of {\.Z}ebru{\'n} \etal (2001b), \ie are 
based on the equatorial coordinates of the star for the epoch J2000 in the 
format:

\centerline{OGLE$hhmmss.ss-ddmmss.s$.} 

\noindent
For example, OGLE053232.28-700056.5 stands 
for a star with coordinates ${\rm RA}=05\uph32\upm32\zdot\ups28$ and 
${\rm DEC=-70\arcd00\arcm56\zdot\arcs5}$. 

1882 stars were classified as EA, 718 as EB and 168 as EW type. These figures 
do not sum up to 2681, because of several double classifications. Appendices 
A--C present examples of DIA {\it I}-band light curves of types EA, EB and EW, 
respectively. The ordinate is the phase with 0.0 value corresponding to the 
deeper eclipse. Abscissa is the $I$-band magnitude. Light curve is repeated 
twice for clarity. 

Tables, light curves and finding charts of all 2681 eclipsing binary objects 
are available from the OGLE {\sc Internet} archive and $via$ the WWW Interface 
(Section~8). 

We should stress that periods of several stars might be incorrect. First, 
period can be two times longer than the real one, because in some cases, the 
secondary eclipse of faint stars and for noisy light curves could not be 
reliably detected. Other possible period error, which was noticed during the 
visual inspection of light curves, was related with the AoV method. For stars 
with large eccentricity it triggered periods of ${1.5\times P}$ or ${2.5\times 
P}$ instead of the correct period $P$. We probably corrected most of such 
cases, but we cannot fully exclude that some of them are still in the catalog. 
\vspace*{-11pt}
\Section{Discussion}
\vspace*{-3pt}
We present 2580 eclipsing binary stars located in the central regions of the 
LMC. It is the largest number of eclipsing stars ever discovered in the LMC. We 
did not attempt to identifyáthe detected stars with previously discovered 
eclipsing binaries in the LMC, but it is obvious, that the vast majority of 
these objects are newly discovered systems, because other published catalogs 
contain only 79 (EROS: Grison \etal 1995) and 611 (MACHO: Alcock \etal 1997) 
objects. 
\begin{figure}[htb]
\centerline{\includegraphics[width=10cm, height=8cm]{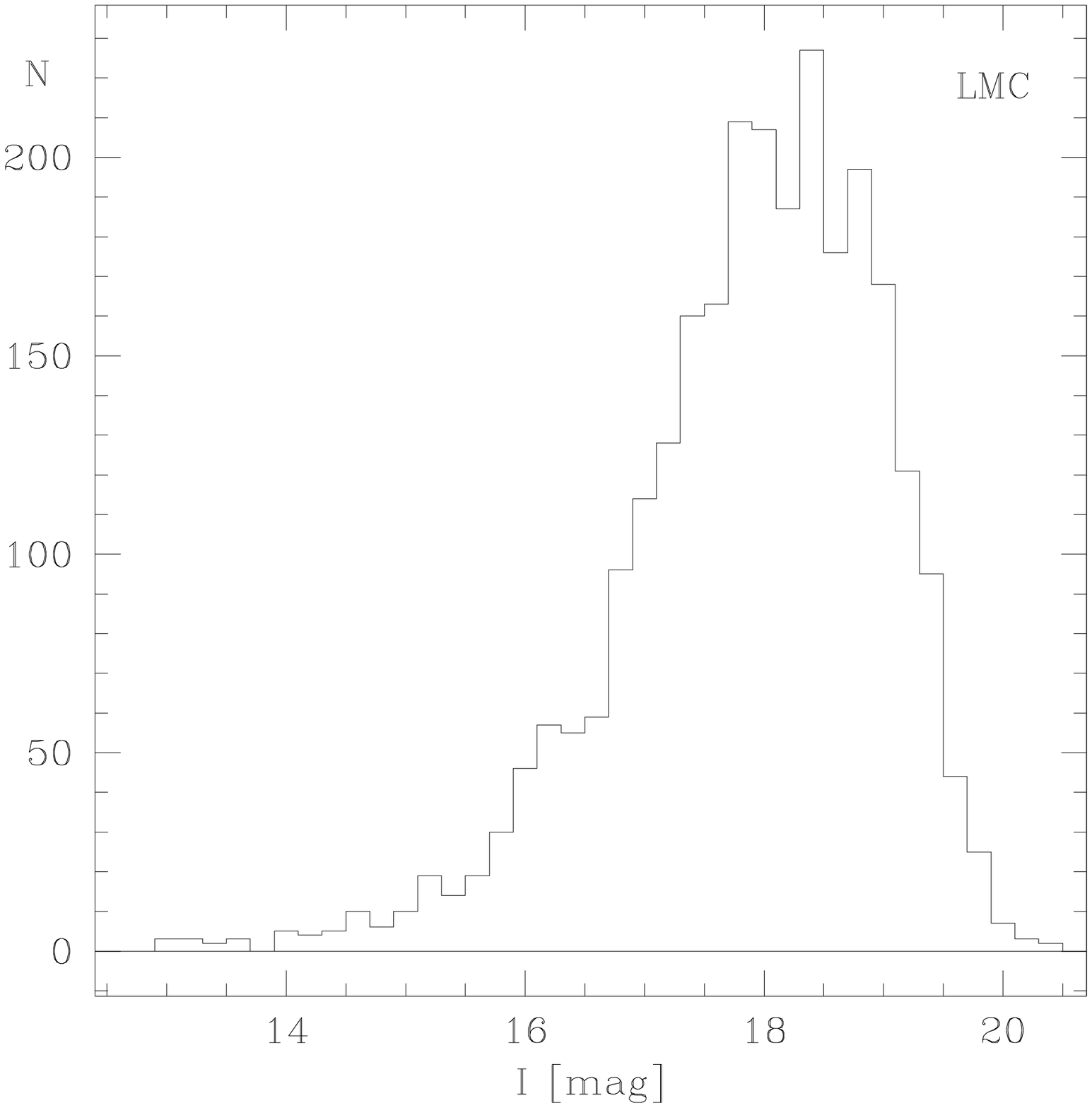}}
\FigCap{Histogram of the DIA $I$-band brightness for eclipsing stars in 0.2~mag 
bins.} 
\end{figure} 

Very good quality of the DIA photometry is clearly visible in the light curves 
of presented stars. Also magnitude limit is low, as shown in Fig.~3, which 
presents the histogram of the DIA $I$-band brightness for all eclipsing stars found 
in the LMC. 

\begin{figure}[htb]
\vspace{5cm}
\centerline{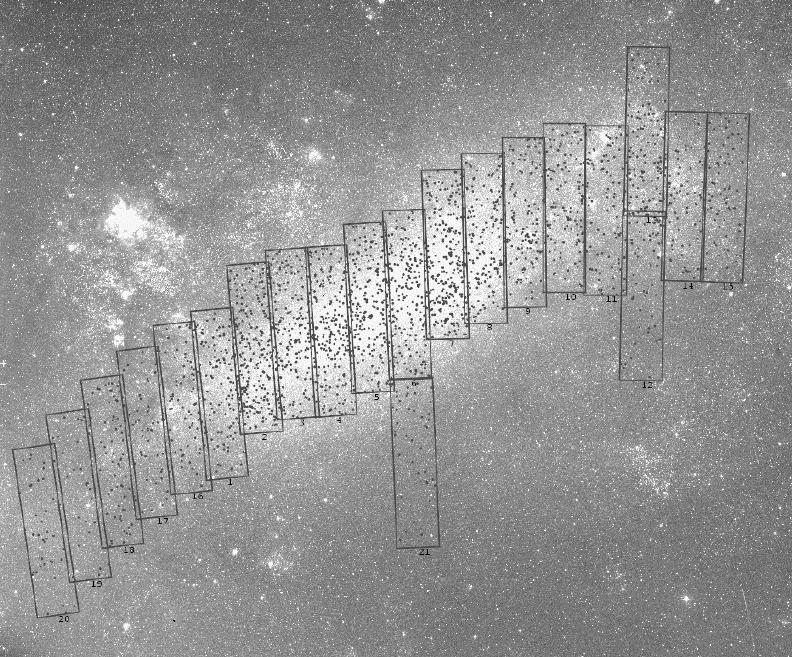}
\vspace{5cm}
\FigCap{OGLE-II fields in the LMC. Dots indicate positions of eclipsing
stars. North is up and East to the left in the DSS image.}
\end{figure}
Fig.~4 presents a picture of the LMC from the Digitized Sky Survey (DSS) with 
contours of the OGLE-II fields. Positions of the eclipsing binary stars are 
marked with black dots. The stars are distributed proportionally to the density 
of the LMC stars, with the largest concentration in the fields 
LMC\_SC2--LMC\_SC7. 

\begin{figure}[htb]
\centerline{\includegraphics[width=11cm, height=8.8cm]{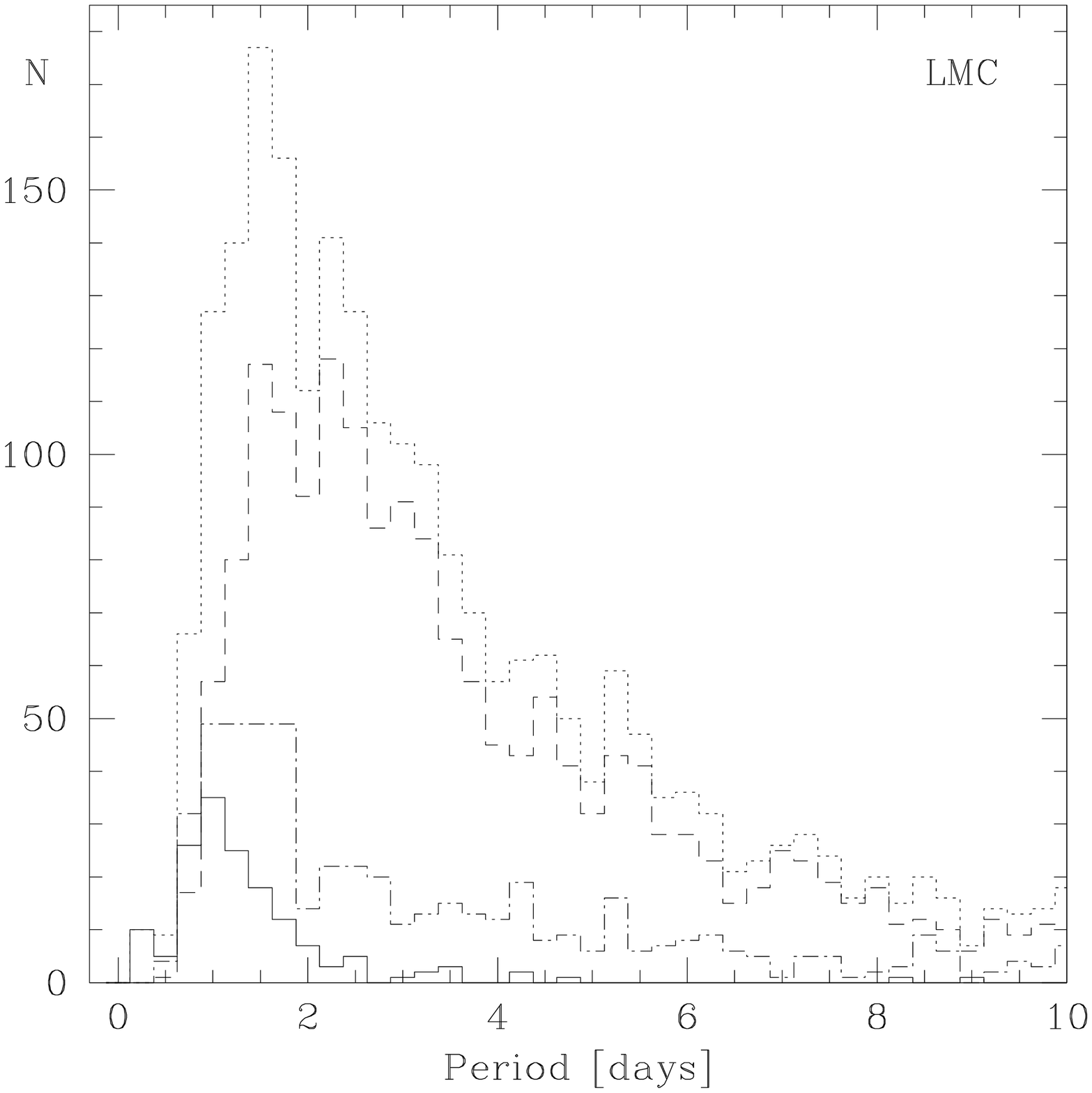}}
\FigCap{Histogram of periods of eclipsing binaries in 0.25~day bins. Dashed, 
dot-dashed and solid lines correspond to classes EA, EB and EW respectively. 
Doted line corresponds to all eclipsing objects. Additional 416 objects have 
periods longer than 10 days.}
\end{figure} 
Fig.~5 shows the histogram of orbital periods of the LMC eclipsing variables in 
0.25 day bins from 0 to 10 days. Dashed, dot-dashed and solid lines correspond 
to classes EA, EB and EW, respectively and dotted line corresponds to all 
eclipsing objects. Additional 416 objects with periods longer than 10 days are 
distributed more or less uniformly and their number falls rapidly to zero at 
longer periods. The longest reliable period found among all objects equals to 
251.096 days (LMC\_SC16 OGLE053725.90-700223.3), but there are two objects 
(LMC\_SC21 OGLE052052.41-700655.1 and LMC\_SC19 OGLE054310.48-703057.8), with 
only one or two eclipses observed, which periods can be even longer. However, 
these periods cannot be derived reliably with the present dataset. 

The majority of stars are short period systems with the most frequent period 
of about 1.5 days. EA type is the most numerous class of eclipsing objects in 
the LMC. Its period distribution is characterized by a broad maximum at 
1.5--2.5 days. Periods of stars from EB class are distributed with wide, flat 
maximum in the range of 1--2 days. The EW class is least numerous and has a 
maximum at periods of about 1~day. Distribution of the latter class is, 
however, severely biased because only long period tail of these objects from 
the LMC is bright enough to be in the range of the OGLE data. The remaining 
part of this class are foreground Galactic variables.

\begin{figure}[htb]
\vglue-10mm
\vspace{6cm}
\centerline{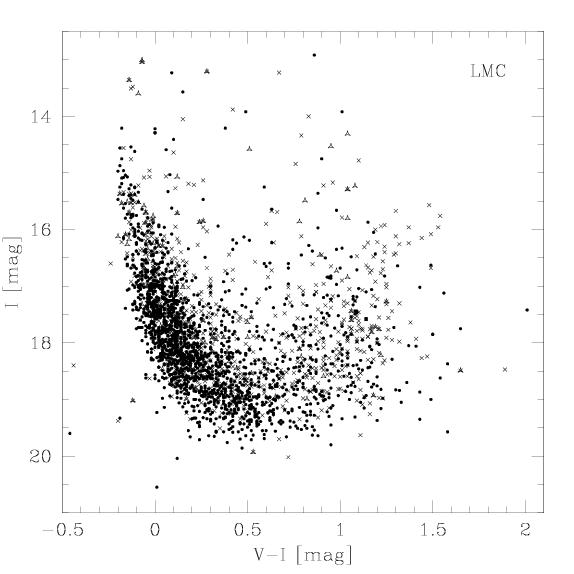}
\vspace{6cm}
\FigCap{Color-magnitude diagram of eclipsing binaries in the LMC. Solid dots, 
crosses and triangles mark EA, EB and EW type objects, respectively.} 
\end{figure} 
Fig.~6. presents {\it I} \vs ${V-I}$ color-magnitude diagram for all eclipsing 
binary stars from the catalog. EA, EB and EW classes are marked with different 
symbols. Fig.~6 indicates, that almost all eclipsing stars belong to the LMC 
and there are only a few foreground systems. All three classes seem to be 
distributed uniformly over the CMD diagram. 

\begin{figure}[htb] 
\vglue-10mm
\centerline{\includegraphics[width=13cm]{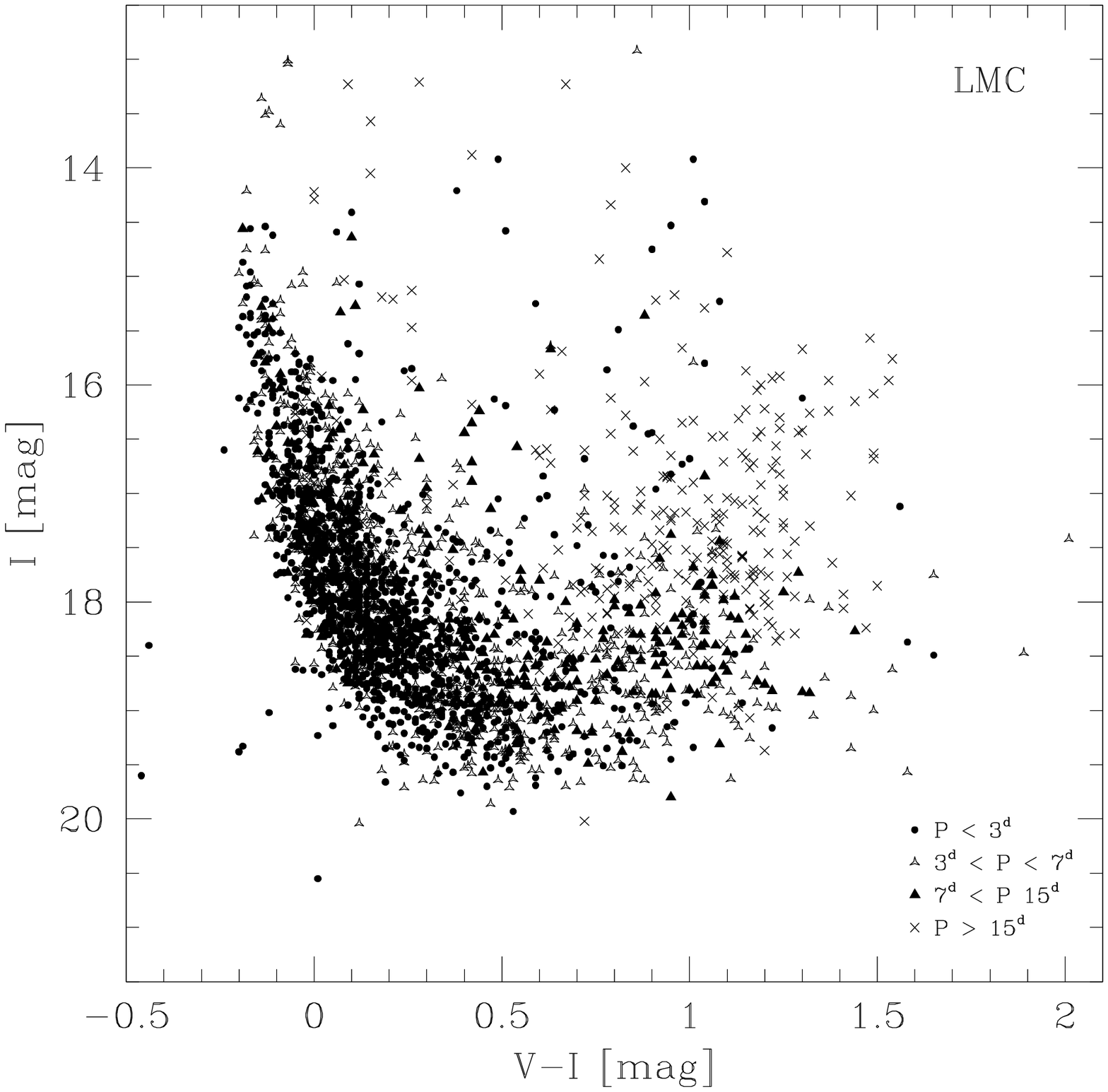}}
\FigCap{Color-magnitude diagram of eclipsing binaries in the LMC, Different 
symbols mark position of stars with short, medium, long and very long periods.} 
\end{figure} 
In Fig.~7 we present CMD diagram for all eclipsing binary stars in the LMC, 
dividing stars into 4 groups depending on their periods: short, medium, long 
and very long. Each group is marked with different symbol. This plot was 
created similarly to the one for the eclipsing binaries found in the SMC (Fig.~3 in 
Udalski \etal 1998a). Both figures indicate close similarities in the 
distribution in the CMD of the eclipsing binary stars from both Magellanic 
Clouds. The majority of short and medium period eclipsing stars belong to the 
young population located on the main sequence. Long period stars are located in 
the lower giant branch or on the right part of the main sequence, \ie they are 
probably evolved main sequence stars. Very long period stars are mostly 
concentrated in the upper part of the red giant branch. 

\Section{Completeness of the Catalog and Network Efficiency}
Assessment of the completeness of our catalog is important for statistical 
studies of eclipsing binaries in the LMC. Fig.~3, showing distribution of 
magnitudes of eclipsing objects, suggests that our sample should be complete to 
about $I\approx17.5$ mag. In general, the completeness is a function of 
magnitude, quality of photometry, period and other factors. We attempted to 
determine the mean completeness of our catalog by comparing objects detected in 
the overlapping regions of neighboring fields. Based on astrometric solutions, 
we checked, which of the detected eclipsing binary stars should have a 
counterpart in the neighboring field and compared these objects with actually 
detected stars. In total, 238 stars should be theoretically paired up. In 
practice 202 stars with pairs were found, yielding the mean completeness of our 
catalog equal to about 85\%. However it should be noted that this is certainly 
a lower limit as the regions close to the edge of each field are affected by 
non-perfect pointing of the telescope leading to effectively smaller number of 
observations. 

Classification types and periods of paired stars were very similar to their 
counterparts, however we unified them to the values of star with larger number 
of observations. 

\begin{figure}[htb]
\centerline{\includegraphics[width=12cm]{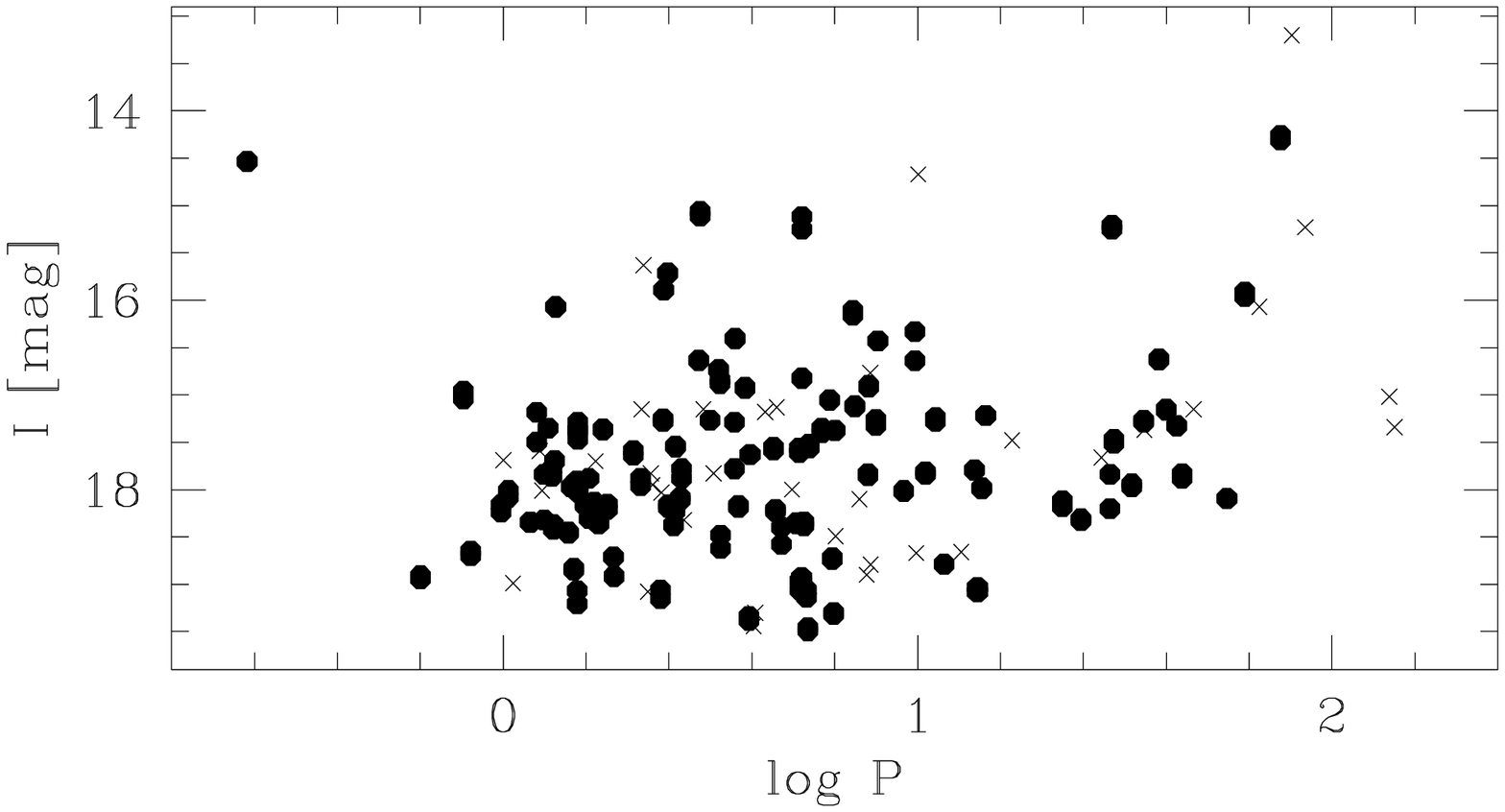}}
\FigCap{Diagram of brightness \vs $\log P$ for eclipsing stars which should be 
paired up with stars located in the neighboring fields. Solid dots mark found 
paired stars and crosses mark those that remained unpaired.} 
\end{figure}
Fig.~8 presents the diagram of brightness of the system \vs $\log P$ for stars 
used in our test. Solid dots mark stars that were paired in the neighboring 
fields. Crosses show missed pairs. It can be seen that completeness is 
depending on brightness -- most unpaired stars are fainter than ${I\approx 
17.5}$~mag. Histogram of brightness of all eclipsing binaries (Fig.~3) also 
indicates, that completeness is likely to be better than 90\% for objects 
brighter than ${I<17.5}$~mag and then it falls gradually to zero for stars 
fainter than ${I\approx20}$~mag. 

We also used paired stars from overlapping fields to check the neural network 
efficiency. Among 238 stars which should be paired, we found 5 different stars 
rejected in our pipeline before entering the net, in most cases because of 
small number of observations. Consequently, only 233 of 238 objects entered the 
net, leading to only 228 to-be-paired objects. Because we found 202 paired 
stars, we checked 26 not paired objects and found that 10 of them were rejected 
by the network (0-class) and one was classified as a ``saw-shape'' object. 
Assuming, that the remaining 15 stars (\ie 6\%) passed correctly through the 
net, but were somehow misclassified or rejected during visual inspection, it 
gives 217 objects, which were correctly classified by the network. Compared to 
228 objects entered, it yields about 95\% neural network efficiency. It is 
worth noticing that the network behavior was ``secure'' in a sense that it 
was rather rejecting eclipsing variables than classifying them to wrong classes 
(one star of 228 gives probability of a mistake less than 1\%). 

Independently, we checked the network efficiency by visual inspection of all 
rejected stars (\ie classified to 0-class) and found additional 146 candidates 
for eclipsing variables. Compared to 2681 eclipsing stars classified correctly, 
it yields again about 94\% network efficiency. However, one should note, that 
these tests are only crude approximation and more advanced tests are needed to 
check neural network efficiency in more detail. Nevertheless, our results are 
very encouraging and confirm good performance of our algorithm based on 
artificial neural network. Therefore, it seems it could be applied successfully 
also in other implementations of automatic classification of variable stars.

\Section{Candidates for Distance Measurements}
Selection of systems suitable for distance determination is one of the main and 
the most important applications of the catalog of eclipsing binary stars. The 
catalog contains large number of bright and well detached systems, suitable for 
spectroscopic study. From EA class, which corresponds to detached eclipsing 
systems, we selected 36 stars, which satisfied the following criteria: 
brightness in the {\it I}-band $<$ 16 mag and depth of secondary minimum $>$ 
0.2~mag. 

\begin{figure}[htb]
\centerline{\includegraphics[width=12.5cm]{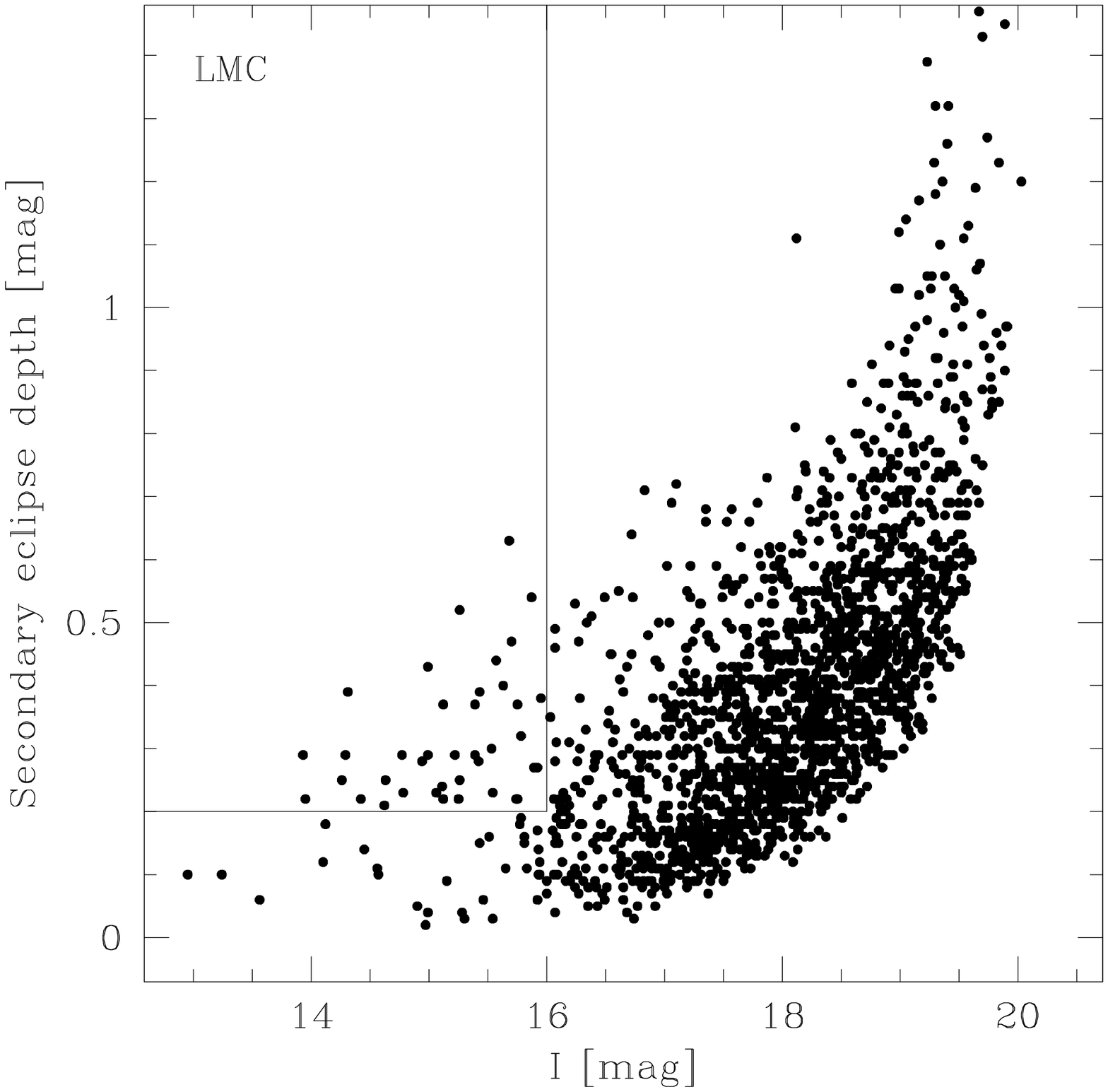}} 
\FigCap{Depth of the secondary eclipse \vs {\it I}-band brightness diagram for EA 
class of eclipsing binaries detected in the LMC. Lines mark the region with 
stars that were selected as candidates for distance measurement.} 
\end{figure} 

\begin{landscape}
\renewcommand{\arraystretch}{1}
\MakeTableSepp{r@{\hspace{11pt}}r@{\hspace{11pt}}r@{\hspace{11pt}}r
@{\hspace{11pt}}r@{\hspace{11pt}}r@{\hspace{11pt}}r
@{\hspace{11pt}}r@{\hspace{11pt}}r@{\hspace{11pt}}r
@{\hspace{11pt}}r}{11cm}{Selected EA eclipsing binary stars suitable for
distance measurements to the LMC}{
\hline
\noalign{\vskip 3pt}
\multicolumn{1}{c}{No.}&
\multicolumn{1}{c}{Field}&
\multicolumn{1}{c}{Star}&
\multicolumn{1}{c}{Period}&
\multicolumn{1}{c}{$T_0$}&
\multicolumn{1}{c}{$I$}&
\multicolumn{1}{c}{$\Delta I_{\rm PRI}$}&
\multicolumn{1}{c}{$\Delta I_{\rm SEC}$}&
\multicolumn{1}{c}{$V-I$}&
\multicolumn{1}{c}{$B-V$}&
\multicolumn{1}{c}{Overlap}\\
&
&
&
\multicolumn{1}{c}{[days]}&
\multicolumn{1}{c}{$-2450000$}&
\multicolumn{1}{c}{(DIA)}&
\multicolumn{1}{c}{(DIA)}&
\multicolumn{1}{c}{(DIA)}&
&
&
\multicolumn{1}{c}{field}\\
\noalign{\vskip 3pt}
\hline
\noalign{\vskip 3pt}
  112&LMC\_SC1&OGLE053459.53-694406.2& 75.262000&564.35923&14.31&0.40&0.39&$ 0.00$&$ -0.19$&LMC\_SC16\\ 
  114&LMC\_SC1&OGLE053502.18-694417.8&  2.989580&454.12044&15.11&0.36&0.24&$-0.17$&$ -0.18$&LMC\_SC16\\
  170&LMC\_SC2&OGLE053022.67-694126.1&  0.645040&455.61994&15.26&0.78&0.25&$ 0.59$&$  0.37$&\\
  196&LMC\_SC2&OGLE053039.28-701409.7& 24.672310&454.56456&14.29&0.31&0.29&$ 0.00$&$ -0.17$&\\
  207&LMC\_SC2&OGLE053046.88-694759.2&  2.917210&453.95612&15.95&1.09&0.38&$-0.04$&$ -0.15$&\\
  314&LMC\_SC2&OGLE053202.60-692717.2& 13.267850&458.92109&15.92&0.29&0.27&$-0.09$&$ -0.22$&\\
  362&LMC\_SC2&OGLE053225.29-692537.4&  3.370250&456.81334&15.26&0.56&0.52&$-0.11$&$ -0.20$&\\
  462&LMC\_SC3&OGLE052835.52-695102.3&  5.638740&457.74472&15.89&1.09&0.27&$-0.01$&$ -0.12$&\\
  630&LMC\_SC4&OGLE052525.66-693304.5&157.554800&380.25443&15.63&0.40&0.30&$ 1.30$&$  1.24$&\\
  824&LMC\_SC5&OGLE052235.46-693143.4&  2.183370&453.36372&15.63&0.43&0.40&$-0.17$&$ -0.17$&\\
  843&LMC\_SC5&OGLE052244.34-693143.5&  2.150540&454.36212&15.54&0.27&0.23&$-0.16$&$ -0.14$&\\
 1129&LMC\_SC6&OGLE052100.81-692944.9&  1.300790&454.76565&15.42&0.29&0.28&$-0.17$&$ -0.15$&\\
 1295&LMC\_SC7&OGLE051750.09-693827.8& 56.234850&448.96592&15.87&0.57&0.54&$ 1.15$&$  1.16$&\\
 1321&LMC\_SC7&OGLE051804.81-694818.9&  3.107040&454.78999&14.77&0.30&0.29&$-0.18$&$ -0.16$&\\
 1368&LMC\_SC7&OGLE051828.18-693745.3&  1.403790&455.42725&15.22&0.30&0.29&$-0.13$&$ -0.16$&\\
 1431&LMC\_SC7&OGLE051858.97-693549.5&  9.143980&461.16709&15.74&0.34&0.22&$-0.15$&$ -0.13$&\\
 1450&LMC\_SC7&OGLE051911.89-694225.0&  2.727130&456.45529&15.43&0.42&0.39&$-0.20$&$-99.99$&\\
 1627&LMC\_SC8&OGLE051644.53-693233.3&  5.603590&457.56235&15.75&0.46&0.37&$-0.13$&$-99.99$&\\
 1716&LMC\_SC9&OGLE051314.85-691432.5&  2.007900&457.58698&13.95&1.58&0.22&$ 0.49$&$  0.33$&\\
 1729&LMC\_SC9&OGLE051323.98-692249.2&  2.636640&456.55989&15.74&0.86&0.82&$ 0.05$&$  0.01$&\\
 1748&LMC\_SC9&OGLE051341.40-693245.5&  5.457280&455.07207&15.53&0.33&0.30&$-0.09$&$ -0.12$&\\
1838&LMC\_SC10&OGLE051019.64-685812.2&214.379070&390.12764&15.70&0.48&0.47&$ 0.98$&$  0.99$&\\
1849&LMC\_SC10&OGLE051028.75-692048.0&  3.773360&457.39415&15.78&0.33&0.32&$-0.12$&$ -0.15$&\\
1870&LMC\_SC10&OGLE051108.69-691216.0&  8.000370&455.95671&15.39&0.38&0.37&$ 0.88$&$  0.70$&\\
1967&LMC\_SC11&OGLE050828.13-684825.1&  2.995410&723.93691&14.63&0.28&0.25&$-0.11$&$ -0.16$&\\
1998&LMC\_SC11&OGLE050929.29-685502.8&  2.678800&727.05172&14.94&0.46&0.28&$-0.17$&$ -0.21$&\\
2001&LMC\_SC11&OGLE050934.33-685425.9&  1.462910&727.75272&15.75&0.23&0.22&$-0.10$&$ -0.14$&\\
2157&LMC\_SC13&OGLE050634.43-682544.2&  2.154480&727.82895&15.12&0.41&0.37&$-0.18$&$ -0.17$&\\
2207&LMC\_SC14&OGLE050234.59-690546.0&  5.255940&725.05728&15.25&0.23&0.22&$-0.15$&$ -0.18$&LMC\_SC15\\
2321&LMC\_SC15&OGLE050140.27-685106.0&  4.034850&725.70581&15.39&0.37&0.29&$-0.09$&$ -0.13$&\\
2366&LMC\_SC16&OGLE053517.75-694318.7&  3.881950&729.45995&14.99&0.47&0.43&$-0.16$&$ -0.16$&\\
2427&LMC\_SC16&OGLE053714.17-702001.5&  3.256680&726.33234&15.68&0.66&0.63&$-0.07$&$ -0.11$&\\
2456&LMC\_SC17&OGLE053818.62-704108.4&  2.191320&724.77328&14.42&0.27&0.22&$ 0.10$&$  0.01$&\\
2535&LMC\_SC18&OGLE054101.94-700504.7&  2.385180&725.42213&15.57&0.47&0.44&$ 0.09$&$  0.02$&\\
2555&LMC\_SC18&OGLE054148.68-703531.0&  2.768950&726.73270&14.62&0.26&0.21&$ 0.06$&$-99.99$&\\
2585&LMC\_SC19&OGLE054324.89-701650.8&  2.054440&725.94139&14.78&0.25&0.23&$ 0.90$&$  0.74$&\\
\hline}
\end{landscape}

Fig.~9 presents the diagram of the {\it I}-band brightness \vs secondary 
eclipse depth. The selection region is marked. Note that for a part of objects 
the depth of the secondary eclipse is deeper than 0.7~mag. For some very 
ellipsoidal stars this can be a real feature caused by ellipsoidal shape and 
gravitational darkening. For the others this is probably an artifact caused by 
crude determination of the depth by automatic procedure in the case of noisy 
photometry, nonaccurate transformation of the DIA flux differences to 
magnitudes in the case of blended objects and uncertain period (${2\times P}$ 
\vs $P$). Table~4 contains main information about the selected stars: number of 
star in the catalog, field, identification, orbital period, heliocentric J.D. 
of the primary minimum (${T_0-2450000}$) corrected for the position of the star 
in driftscan image, {\it I}-band magnitude, {\it I}-band amplitude (\ie depth 
of the primary minimum), {\it I}-band depth of the secondary minimum, ${V-I}$ 
color, ${B-V}$ color and overlapping field if a given star is also present. 
Color value of ${-99.99}$ stands for no observations in the {\it B} or {\it V} 
bands. Appendix~D presents the DIA {\it I}-band light curves of 36 selected 
stars. The ordinate is phase with 0.0 value corresponding to the deeper 
eclipse. Abscissa is {\it I}-band magnitude. Magnitudes at the top and bottom 
left are the brightness of the maximum and primary minimum, respectively. 
Period of the star is also given. Light curve is repeated twice for clarity.

All information about selected eclipsing binaries, light curves as well as 
finding charts can be found in the OGLE {\sc Internet} archive and {\it via} WWW 
Interface (see below). 

\Section{The Catalog in the I{\small NTERNET}}
The Catalog of eclipsing binary stars is available on-line through {\sc ftp} 
and WWW from the OGLE {\sc Internet} archive. The Catalog can be accessed {\it 
via} anonymous {\sc ftp} at the following addresses: 
\vskip3pt
\centerline{\it ftp://sirius.astrouw.edu.pl/\~{}ogle/ogle2/var\_stars/lmc/ecl}
\centerline{\it ftp://bulge.princeton.edu/\~{}ogle/ogle2/var\_stars/lmc/ecl}
\vskip3pt
\noindent
WWW interface to the catalog is available from the following addresses:
\vskip3pt
\centerline{\it http://www.astrouw.edu.pl/\~{}ogle}
\centerline{\it http://bulge.princeton.edu/\~{}ogle}

The Catalog will be regularly updated when the final set of the OGLE-II data 
is available and/or any errors, unavoidable in so large dataset, are found. 
The most recent version will always be available in the {\sc Internet} from 
the above addresses. The Catalog will also be significantly extended when 
large enough number of epochs in the ongoing OGLE-III phase is collected. As 
the OGLE-III fields cover practically entire LMC the final version of the 
Catalog will include the vast majority of eclipsing stars from the LMC. 

\Section{Summary}
The catalog of eclipsing binary stars found during the OGLE-II project in the 
Large Magellanic Cloud is the largest set of such type variable stars. The 
catalog contains 2580 eclipsing stars of three main types EA, EB and EW. Its 
high completeness which was obtained by using automated search algorithm based 
on artificial neural network makes it very useful for many statistical 
analysis of the LMC stars. Very good quality of photometry and very long 
time-base of the OGLE-II observations allowed to obtain good quality light 
curves for most objects in the catalog and very accurate orbital periods. 
Additionally we selected a subsample of bright detached eclipsing systems 
suitable for distance determinations. With the high resolution spectra 
obtained with the largest 6-8 m class telescopes very accurate distance 
determinations to these objects should be potentially possible, allowing 
independent, accurate determination of the distance to the LMC. 

\Acknow{We would like to thank Prof.~Bohdan Paczy{\'n}ski for his 
encouragements and discussions about this work. We would also like to thank 
Dr.~S{\l}avek Ruci{\'n}ski for his help, and Dr.~Grzegorz Pojma{\'n}ski and 
Tomasz Mizerski for making their computer programs available. This work was 
partly supported by the KBN grant 2P03D02523 to {\L}.~Wyrzykowski and NASA 
grant NAG5-12212 and NSF grant AST-0204908 to B.~Paczy\'nski. We acknowledge usage 
of the Digitized Sky Survey which was produced at the Space Telescope Science 
Institute based on photographic data obtained using the UK Schmidt Telescope, 
operated by the Royal Observatory Edinburgh.}

\begin{figure}[p]
\centerline{\large\bf Appendix A}
\vskip9pt
\centerline{\bf Eclipsing stars in the LMC}
\vskip5pt
\centerline{\bf EA type eclipsing stars}

\hglue-3mm{see OGLE {\sc Internet} Archive}
\end{figure} 
\begin{figure}[p]
\centerline{\large\bf Appendix B}
\vskip9pt
\centerline{\bf Eclipsing stars in the LMC}
\vskip5pt
\centerline{\bf EB type eclipsing stars}

\hglue-3mm{see OGLE {\sc Internet} Archive}
\end{figure} 
\begin{figure}[p]
\centerline{\large\bf Appendix C}
\vskip9pt
\centerline{\bf Eclipsing stars in the LMC}
\vskip5pt
\centerline{\bf EW type eclipsing stars}

\hglue-3mm{see OGLE {\sc Internet} Archive}
\end{figure} 
\begin{figure}[p]
\centerline{\large\bf Appendix D}
\vskip9pt
\centerline{\bf Eclipsing stars in the LMC}
\vskip5pt
\centerline{\bf Selected detached eclipsing stars}

\hglue-3mm{files: 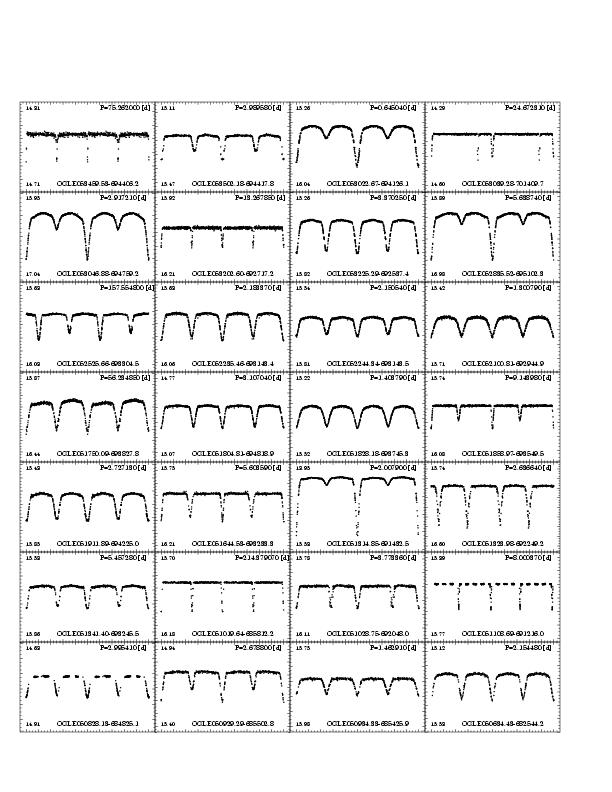, 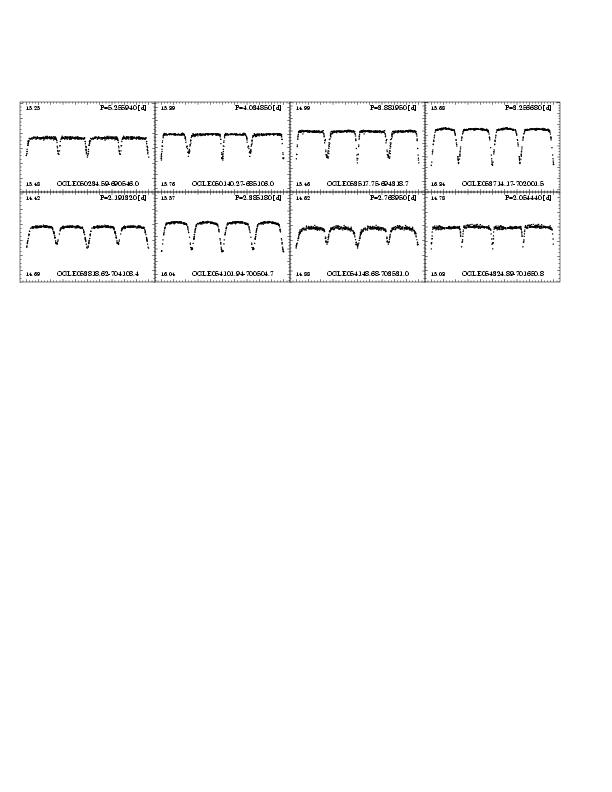}
\end{figure} 

\end{document}